\documentclass[a4paper, 12pt]{article}
\usepackage{xcolor}
\usepackage[utf8]{inputenc}
\usepackage{authblk}
\usepackage{amsmath}
\usepackage{amssymb}
\usepackage{natbib}
\usepackage{graphicx}
\title{\textbf{Solitary waves of moderate amplitude and dispersive radiation in
the Serre equations: the extended KdV–Whitham approximation}}
\author[1]{Benjamin Martin}
\author[1]{Dmitri Tseluiko}
\author[1]{Karima Khusnutdinova \footnote{Corresponding author: K.Khusnutdinova@lboro.ac.uk}}
\affil[1]{\small Department of Mathematical Sciences, Loughborough University, LE11 3TU, United Kingdom}
\date{}

\begin{document}

\newcommand{\crr}{\color{black}}
\newcommand{\crb}{\color{blue}}
\newcommand{\crg}{\color{green}}

\maketitle


\abstract{
We consider the extended Korteweg–de Vries (eKdV) equation as a model for long moderately nonlinear surface water waves {\crr and use it to describe the evolution of initial conditions generating solitary waves with and without significant dispersive radiation, as well as cases of pure dispersive radiation without any solitary waves.}  In the slow time formulation {\crr for the modelled solutions}  this equation {\crr also} generates fast propagating resonant {\crr forward} radiation due to the non-convexity of its linear dispersion curve, which is not present in the {\crr direct numerical simulations of the} strongly nonlinear Serre parent system {\crr (also known as the Su-Gardner and Green-Naghdi equations)}. We show that the extended KdV–Whitham approximation and the slow space formulation of the eKdV equation are suitable regularisations of the eKdV equation in several cases of interest. {\crr Importantly, unlike the KdV-type equations, it can be used to model waves of} moderate amplitude. Numerical comparisons are made between the {\crr Serre} system and {\crr several} respective reduced models, where simulations are initiated with an approximate soliton solution of the eKdV equation, constructed by use of Kodama-Fokas-Liu near-identity transformation to the KdV equation, {\crr as well as a generic localised initial condition}.
}

\section{Introduction}
\label{section_intro}

{\crr Long weakly nonlinear surface} waves have been studied at length and relevant reduced equations and their approximate solutions have been  {\crr considered and tested} for numerous contexts (see, for example, \citet{W74, AB81, D89, J97, MKD15, TKCO16, KLPS18, HFMS21, CHP24, MTK25} and the references therein). The most well known model is the Korteweg--de Vries (KdV) equation given by
\begin{equation}
\eta_T + \alpha \eta \eta_{\xi} + \beta \eta_{\xi\xi\xi} = 0,
\end{equation}
where $\eta$ is the wave amplitude, $T = \epsilon t$ is a slow time variable given the small amplitude parameter $\epsilon$, $\xi = x - c_0 t$ is a fast spacial variable in a coordinate frame moving with the linear long wave speed $c_0$, and $\alpha$, $\beta$ are some context dependent coefficients. In this study of right-propagating surface water waves the non-dimensional speed $c_0 = 1$. The fundamental balance between nonlinearity and dispersion allows the KdV equation to accurately describe many solutions, including solitons and {\crr dispersive shock waves}. This equation is not constrained to surface water waves, it has been used {\crr with the necessary adjustments}  in  many different physical contexts such as, for example, internal waves \citep{OS80}, waves with a shear flow \citep{G05}, waves over bottom topography \citep{J73}, and waves in solids \citep{KS08} to name but a few. It is well known that the KdV equation  possesses an infinite number of conservation laws and the initial value problem given {\crr suitable}  Cauchy data can be solved analytically by the Inverse Scattering Transform (IST) \citep{GGKM67, GGKM74}.

However, despite this universality the KdV equation is valid in the asymptotic limit as the small amplitude parameter $\epsilon \rightarrow 0$ describing weakly nonlinear, weakly dispersive waves. As $\epsilon$ increases the physical wave amplitude increases, so too the nonlinearity, and the accuracy of the KdV equation deteriorates. The KdV equation also struggles to accurately describe {\crr dispersive} wave trains due to being a long wave model, i.e. only possessing leading order dispersion accuracy. It is therefore natural to include the second order of the asymptotic expansion used in the derivation to extend the region of validity to moderately nonlinear {\crr dispersive} waves. Doing so yields the extended KdV (eKdV) equation written as
\begin{equation}
\eta_T + \alpha \eta \eta_{\xi} + \beta \eta_{\xi\xi\xi} + \epsilon \left( \alpha_1 \eta^2 \eta_{\xi} + \gamma_1 \eta \eta_{\xi\xi\xi} + \gamma_2 \eta_{\xi} \eta_{\xi\xi} + \beta_1 \eta_{\xi\xi\xi\xi\xi} \right) = 0 \label{1}
\end{equation}
{\crr (see \citet{B66, W74, MS90} and references therein).} The small amplitude parameter is present explicitly in this model.
The eKdV equation has similarly been derived and used in many different physical contexts, for example, surface water waves \citep{HFMS21,HFS22}, internal water waves \citep{LY96,STCK25}, gravity-capillary waves \citep{AS85,HS88,B91}, plasma physics \citep{KO69}, waves under an ice sheet \citep{GP14} and undular bores generated by fracture in solids \citep{HRHK21}. There also exist counterparts in cylindrical geometry \citep{HFMS21, STCK24} that were shown to perform considerably better than the respective leading order approximation of the {\crr cylindrical KdV} equation for moderate amplitudes, see \cite{STCK24}. There are fewer analytical results for the eKdV equation as it is not integrable, except for  special parameter values, which are not the subject of this study, where we concern ourselves with the generic case.

Rather than {\crr retaining} all of the nonlinear terms that appear at the second order it can be reasoned that when the coefficient of the leading order quadratic nonlinearity term is small or vanishing it could be useful to include only the next nonlinear term in the asymptotic expansion. This can be obtained from the eKdV equation by setting $\gamma_1 = \gamma_2 = \beta_1 = 0$ and this yields the truncated Gardner equation
\begin{equation}
\eta_T + \alpha \eta \eta_{\xi} + \beta \eta_{\xi\xi\xi} + \epsilon \alpha_1 \eta^2 \eta_{\xi} = 0. \label{tru_Gardner}
\end{equation}
{\crr The Gardner equation admits soliton solutions, including table-top solitons. It has been used to model many nonlinear wave processes observed in the oceans } (see \citet{GPT99, GPPS02, G05, KRTKP16} and the references therein). Recently, it has been shown that an improvement can be made to the coefficient of the cubic nonlinearity term by applying a nonlocal {\crr Kodama-Fokas-Liu} near-identity transformation (NIT)  \citep{K85a,K85b,FL96} to the full eKdV equation (\ref{1}) \citep{GBK20, STCK24}. The result is an `improved  Gardner equation' such that
\begin{equation}
B_T + \alpha BB_{\xi} + \beta B_{\xi\xi\xi} + \epsilon \alpha_2 B^2 B_{\xi} = 0, \label{imp_Gardner}
\end{equation} 
which retains only cubic nonlinearity at the second order and $\alpha_2 = (18 \alpha_1 \beta^2 - 2 \beta_1 \alpha^2 - 3\beta \gamma_1 \alpha) / 18\beta^2$. 
In our study we shall compare both numerical and analytical solutions of such Gardner-type equations to both the eKdV equation and the full parent system.

Due to the higher order accuracy of both nonlinear and dispersive terms of the eKdV equation, it has been useful in describing {\crr dispersive shock waves}, see for example \citet{HRHK21}. However, in the context of surface water waves in the slow time formulation  the eKdV equation exhibits resonant radiation that appears in front of the main wave. This has been studied numerically and analytically with descriptions provided for a soliton \citep{BGK93,HFMS21}, {\crr dispersive shock waves} in the eKdV equation \citep{BS23,BHF25}, and there are similar studies for {\crr dispersive shock waves} in the Kawahara equation \citep{SH17,HSS19}. Qualitatively similar formations have also been seen for solitons in the nonlinear Schr\"{o}dinger equation \citep{AKM96}. This behaviour is explained by the linear dispersion relation being non-convex allowing for resonance between waves with  large and small wavenumbers. In what follows we show that the dispersion relation of the slow space version of the eKdV equation and parent 1D {\crr Serre} system \citep{S53,SG69, GN76} {\crr does not have this feature}, and  resonance is not possible. Recently it was shown by \citet{HPS25} {\crr in a study related} to the {\crr cylindrical KdV} equation that if the slow time variation is well posed then the slow space variation is ill-posed and vice-versa. {\crr Also,} no resonant radiation was seen in the slow space extended {\crr cylindrical KdV} equation results of \citet{STCK24} but was seen in the slow time formulations of \citet{HFMS21}. It would therefore be beneficial to regularise the eKdV equation in an efficient manner for cases where the slow time formulation is preferable allowing one to use one and the same initial condition for both the reduced amplitude equations and the parent model. The KdV-Whitham (KdVW) approximation \citep{W67,FW78,KLPS18} replaces the dispersion relation of the KdV equation with that of the parent system and has been shown to perform excellently for {\crr dispersive shock waves}, see \citet{LY96,TKCO16,C18,CHP24} and the references therein. {\crr There has also been recent studies devoted to the Gardner--Whitham-type equations \citep{FP25}.} We show that it is beneficial to introduce the Whitham approximation to regularise the eKdV equation, which we call the extended KdV--Whitham (eKdVW) approximation, and demonstrate the range of its validity. {\crr We also include comparisons with the Benjamin--Bona-Mahoney (BBM) equation \citep{BBM72}, in the fixed reference frame given by
\begin{equation}
{\crr \eta_t + \eta_x +  \epsilon (\alpha \eta\eta_{x} - \beta \eta_{xxt}) = 0.}
\label{BBM}
\end{equation}
The BBM equation regularises high wavenumber shortcomings of the KdV equation by altering the dispersive term to include mixed time and space derivatives. This regularisation is local compared to the KdVW and eKdVW equations where the `correct' linear dispersion is obtained at the expense of a nonlocal pseudo-differential operator.}

The contents of this paper are as follows. In section \ref{section_derivs} we derive the slow time and slow space versions of the KdV and eKdV equations from the fully nonlinear and dispersive {\crr Serre} equations. In section \ref{section_disp_rel} we discuss the linear dispersion relations of the KdV, eKdV, {\crr and BBM} equations compared to the parent {\crr  Serre} equations highlighting the inaccuracy for large wavenumbers that the reduced models posses and why there is resonance in the slow time eKdV equation. We also introduce the eKdVW approximation. In section \ref{section_solitary_sols} we discuss an asymptotic approximation of a solitary wave solution to the eKdV equation using a Kodama-Fokas-Liu NIT to the KdV equation. This yields a solitary wave solution accurate to $O(\epsilon^2)$ which we show to work better  for surface waves compared to the NIT solution constructed from the improved Gardner equation when compared with the {\crr Serre equations} soliton solution. In section \ref{section_solitary_num} we demonstrate numerically the accuracy of the relevant reduced order models comparing the KdV, eKdV, both truncated and improved Gardner equations, {\crr the BBM equation}, and the eKdVW equation for several scenarios and demonstrate that it is possible to predict which method to choose by considering the conserved quantities of the KdV equation and using the IST. In section \ref{section_conclusion} we conclude on our findings, while in appendix \ref{appendixA} we derive the eKdV equation from the Boussinesq-Peregrine  equations, and in appendix \ref{appendixB} we discuss  the numerical methods.

\section{Derivation of the extended KdV equation}
\label{section_derivs}

In this section we briefly derive the KdV and eKdV equations in the slow time and slow space variables from the {\crr Serre} equations for flat bottom topography \citep{S53,SG69,GN76} using asymptotic multiple-scale expansions. The  {\crr Serre} equations are widely accepted as an accurate asymptotic approximation of the Euler equations for {\crr long} surface gravity waves {\crr (e.g. \citet{DCMM13,MDC17, L20, KS21, GK24} and references therein)} and are derived with no assumption on the smallness of the amplitude.
They also admit an exact soliton solution and Hamiltonian form, therefore serving as an accurate benchmark to test and derive reduced order models from in the moderately nonlinear regime.

We first present the 1D  {\crr non-dimensional Serre} equations written as
\begin{align}
&\eta_t + \left[ (1 + \epsilon \eta) u \right]_x = 0, \label{SGN1D1} \\
&u_t + \epsilon uu_x + \eta_x - \frac{\delta^2}{3(1 + \epsilon \eta)} \left( (1 + \epsilon \eta)^3 (u_{tx} + \epsilon uu_{xx} - \epsilon u_x^2) \right)_x = 0, \label{SGN1D2}
\end{align}
where $u$ is the depth averaged horizontal velocity in the $x$-direction, $\eta$ is the free surface elevation, $\epsilon = O(1)$ is the amplitude parameter, and $\delta \ll 1$ is the wavelength parameter. Assuming that $\epsilon \ll 1$, $\epsilon = O(\delta^2)$ and  retaining up to $O(\epsilon)$ terms one obtains the 1D Boussinesq--Peregrine (BP) equations (see \citet{P67} and the references therein) as
\begin{align}
&\eta_t + \left[ (1 + \epsilon \eta) u \right]_x = 0, \label{Bouss1D1} \\
&u_t + \epsilon uu_x + \eta_x - \frac{\delta^2}{3} u_{txx} = 0. \label{Bouss1D2}
\end{align}
A derivation from this model is presented in appendix \ref{appendixA}. Continuing with the full equations and applying the change of variables $(t,x) \rightarrow (T,\xi)$ to (\ref{SGN1D1}) and (\ref{SGN1D2}), where $T = \epsilon t$ is the slow time variable and $\xi = x-t$ is the fast spacial variable in the moving coordinate frame, we obtain
\begin{align}
&\epsilon \eta_T + [(1 + \epsilon \eta)u - \eta]_{\xi} = 0, \label{sgntxi1} \\
&\epsilon u_T - u_{\xi} + \epsilon uu_{\xi}
 + \eta_{\xi} - \frac{\epsilon}{3(1 + \epsilon \eta)} \left[(1 + \epsilon \eta)^3 (\epsilon u_{T\xi} - u_{\xi\xi} + \epsilon uu_{\xi} - \epsilon u_{\xi}^2) \right]_{\xi} = 0. \label{sgntxi2}
\end{align}
We seek a solution of (\ref{sgntxi1}) and (\ref{sgntxi2}) in the form of the asymptotic multiple scale expansion
\begin{equation}
\eta(T,\xi) = \eta^{(0)}(T,\xi) + \epsilon \eta^{(1)}(T,\xi) + \epsilon^2 \eta^{(2)}(T,\xi) + O(\epsilon^3),
\end{equation}
and similarly for $u$. To leading order, $O(1)$, we obtain
\begin{equation}
\eta^{(0)}_{\xi} = u^{(0)}_{\xi},
\end{equation}
and hence if the wave propagates into an unperturbed medium, $\eta^{(0)} = u^{(0)}$. At the next order, $O(\epsilon)$, we obtain 
\begin{align}
&\eta^{(1)}_{\xi} - u^{(1)}_{\xi} = \eta^{(0)}_T + (u^{(0)}\eta^{(0)})_{\xi}, \label{sgnoe1} \\
&u^{(1)}_{\xi} - \eta^{(1)}_{\xi} = u^{(0)}_T + u^{(0)}u^{(0)}_{\xi} + \frac 13 u^{(0)}_{\xi\xi\xi}, \label{sgnoe2}
\end{align}
from which taking the sum of (\ref{sgnoe1}) and (\ref{sgnoe2}), and substituting the leading order relation $u^{(0)} = \eta^{(0)}$, yields the familiar KdV equation
\begin{equation}
\eta_T^{(0)} + \frac 32 \eta^{(0)}\eta^{(0)}_{\xi} + \frac 16 \eta^{(0)}_{\xi\xi\xi} = 0. \label{KdV}
\end{equation}
Equation (\ref{KdV}) is the well known leading order approximation of weakly nonlinear {\crr weakly dispersive} surface water waves. At the third order, $O(\epsilon^2)$, we obtain
\begin{align}
&\eta^{(2)}_{\xi} - u^{(2)}_{\xi} = \eta^{(1)}_T + (u^{(0)}\eta^{(1)})_{\xi} + (u^{(1)}\eta^{(0)})_{\xi}, \label{sgnoes1} \\
&u^{(2)}_{\xi} - \eta^{(2)}_{\xi} = u^{(1)}_T + (u^{(0)}u^{(1)})_{\xi} + \frac 13 (u^{(0)}u^{(0)}_{\xi\xi})_{\xi} + \eta^{(0)}u^{(0)}_{\xi\xi} + \frac 23 \eta^{(0)}u^{(0)}_{\xi\xi\xi} + \frac 13 u^{(1)}_{\xi\xi\xi} - \frac 13 u^{(0)}_{T\xi\xi}. \label{sgnoes2}
\end{align}
Before proceeding it is convenient to determine relations for $O(1)$ and $O(\epsilon)$ terms to simplify the expressions given by  (\ref{sgnoes1}) and (\ref{sgnoes2}). From the KdV equation (\ref{KdV}) and the second order expression given by (\ref{sgnoe1}) we obtain the relations
\begin{align}
&\eta^{(0)}_T = - \frac 32 \eta^{(0)}\eta^{(0)}_{\xi} - \frac 16 \eta^{(0)}_{\xi\xi\xi}, \label{expression1}  \\
&\eta^{(0)}_{T\xi\xi} = - \frac 92 \eta^{(0)}_{\xi} \eta^{(0)}_{\xi\xi} - \frac 32 \eta^{(0)}\eta^{(0)}_{\xi\xi\xi} - \frac 16 \eta^{(0)}_{\xi\xi\xi\xi}, \label{expression2} \\
&u^{(1)} = \eta^{(1)} - \frac 14 \left. \eta^{(0)} \right.^2 + \frac 16 \eta^{(0)}_{\xi\xi}. \label{expression3}
\end{align}
Taking the sum of (\ref{sgnoes1}) and (\ref{sgnoes2}), and substituting (\ref{expression1}), (\ref{expression2}), and (\ref{expression3}), we remove the dependence on the velocity to obtain 
\begin{equation}
2 \eta_T^{(1)} + 3 \left( \eta^{(0)} \eta^{(1)} \right)_{\xi} + \frac 13 \eta^{(1)}_{\xi\xi\xi} - \frac{3}{4} \left. \eta^{(0)} \right.^2 \eta^{(0)}_{\xi} + \frac{23}{12} \eta^{(0)}_{\xi} \eta^{(0)}_{\xi\xi} + \frac 56 \eta^{(0)} \eta^{(0)}_{\xi\xi\xi} + \frac{1}{12} \eta^{(0)}_{\xi\xi\xi\xi\xi} = 0. \label{nearlyeKdV}
\end{equation}
Since the aim is to approximate $\eta$ in the asymptotic expansion we take the approximation to be $\eta \simeq \eta^{(0)} + \epsilon \eta^{(1)}$. Taking the sum of the KdV equation (\ref{KdV}) and $\epsilon$ lots of (\ref{nearlyeKdV}) yields the eKdV equation in $\eta(T,\xi)$ variables as
\begin{equation}
\eta_T + \alpha \eta \eta_{\xi} + \beta \eta_{\xi\xi\xi} + \epsilon \left( \alpha_1 \eta^2 \eta_{\xi} + \gamma_1 \eta \eta_{\xi\xi\xi} + \gamma_2 \eta_{\xi} \eta_{\xi\xi} + \beta_1 \eta_{\xi\xi\xi\xi\xi} \right) = 0, \label{eKdV}
\end{equation}
where we have truncated higher order terms of $O(\epsilon^2)$ and the coefficients $(\alpha, \beta, \alpha_1, \gamma_1, \gamma_2, \beta_1 )$ take the values $(3/2, 1/6, -3/8, 5/12, 23/24, 1/24)$. Importantly for initiating numerical simulations of the 1D {\crr Serre} equations one should use consistent orders for both $\eta$ and the Cartesian velocity $u$. We therefore obtain the same order approximation for $u \simeq u^{(0)} + \epsilon u^{(1)}$ from (\ref{expression3}) to be
\begin{equation}
u = \eta + \epsilon \left( \frac 16 \eta_{\xi\xi} - \frac 14 \eta^2 \right). \label{u}
\end{equation}
This will provide better accuracy when comparing numerical solutions between the parent and reduced order models.

It is not necessary to repeat the derivation for the slow space evolution variable, $X = \epsilon x$. Instead we perform the change of variables $X = \epsilon \xi + T$ to (\ref{eKdV}) and truncate $O(\epsilon^2)$ terms, from which we obtain
\begin{equation}
\eta_X + \alpha \eta \eta_{\xi} + \beta \eta_{\xi\xi\xi} + \epsilon \left( \alpha_1 \eta^2 \eta_{\xi} + \gamma_1 \eta \eta_{\xi\xi\xi} + \gamma_2 \eta_{\xi} \eta_{\xi\xi} + \beta_1 \eta_{\xi\xi\xi\xi\xi} \right) = 0, \label{eKdVX}
\end{equation}
where now $(\alpha, \beta, \alpha_1, \gamma_1, \gamma_2, \beta_1 )$ take the values seen in table \ref{table_ekdv_coeffs}. In appendix \ref{appendixA} we perform the derivation of the eKdV equation from the 1D BP equations (\ref{Bouss1D1}) and (\ref{Bouss1D2}), where the resulting coefficients can be seen in table \ref{table_ekdv_coeffs} too.
\begin{center}
\begin{table*}[ht]
\begin{tabular*}{\textwidth}{@{\extracolsep\fill}lllllllll@{}}
\hline\\[-0.5ex]
\textbf{Parent System} & \textbf{Variables}  & \textbf{$\alpha$} & \textbf{$\beta$} & \textbf{$\alpha_1$}  & {\textbf{$\gamma_1$}}  & \textbf{$\gamma_2$} & \textbf{$\beta_1$} & \textbf{$\alpha_2$}  \\[1.5ex]
\hline\\[-0.5ex]
1D {\crr Serre} & $T, \xi$ & $\frac 32$ & $\frac 16$ & $-\frac 38$     & $\frac{5}{12}$  & $\frac{23}{24}$  & $\frac{1}{24}$  & $-\frac{11}{8}$ \\[1.5ex]
1D {\crr Serre} & $X, \xi$ & $\frac 32$ & $\frac 16$ & $-\frac{21}{8}$ & $-\frac{7}{12}$ & $-\frac{31}{24}$ & $-\frac{1}{24}$  & $-\frac{11}{8}$ \\[1.5ex]
1D BP  & $T, \xi$ & $\frac 32$ & $\frac 16$ & $-\frac 38$     & $\frac{1}{4}$   & $\frac{7}{24}$   & $\frac{1}{24}$  & $-\frac{9}{8}$ \\[1.5ex]
1D BP  & $X, \xi$ & $\frac 32$ & $\frac 16$ & $-\frac{21}{8}$ & $-\frac{3}{4}$  & $-\frac{47}{24}$ & $-\frac{1}{24}$  & $-\frac{9}{8}$ \\[1.5ex]
\hline
\end{tabular*}
\caption{Coefficients of the eKdV equation derived from the respective parent systems and variables, and the {\crr coefficient $\alpha_2$} of cubic nonlinearity term in the improved Gardner equation.}
\label{table_ekdv_coeffs}
\end{table*}
\end{center}
The coefficients presented in table \ref{table_ekdv_coeffs} are consistent with the {\crr far-field limit ($T \rightarrow \infty$ or $R \rightarrow \infty$, depending on the version used)} of the extended {\crr cylindrical KdV} equation derived by \citet{STCK24}. However, they differ from the derivations for example of \citet{HFMS21,HFS22} since {\crr the parent models used to derive the eKdV equation were different}.

\section{Linear dispersion relations}
\label{section_disp_rel}

In this section we discuss the linear dispersion relations of the KdV, eKdV, {\crr and BBM} equations, and the parent {\crr Serre} equations. The models in this paper form four groups of linear dispersion relations. The first is the KdV and Gardner equations, second is the eKdV equation derived from both the {\crr Boussinesq-Peregrine} and {\crr Serre} equations, the third is the BBM equation, {\crr and the fourth is the} 1D {\crr Serre} equations. Hence it is sufficient to compare only one equation from each group.

Resonant radiation is well understood to manifest in front of the main wave feature in the eKdV equation, see for example \citet{HFMS21,BS23} and the references therein. This is due to the linear dispersion relation of the eKdV equation, for the context of surface water waves, being non-convex allowing waves with high wavenumbers to be in resonance with waves with small wavenumbers. This feature is useful in the context of gravity capillary waves and the Kawahara equation where surface tension is important and the third order dispersive term is close to zero \citep{SH17,HSS19}. Modelling long surface water waves where surface tension is neglected we do not obtain this behaviour from the parent models.

The dispersion relation of the KdV and eKdV equations is found by substituting $\eta = e^{i(k\xi - \omega T)}$ into the linearised KdV and eKdV equations (\ref{KdV}) and (\ref{eKdV}). {\crr For the BBM equation (\ref{BBM}) we linearise and substitute $\eta = e^{i(kx - (k + \epsilon \omega )t)}$.} After simplification we obtain the linear dispersion relations 
\begin{align}
 \text{KdV}: &\qquad \omega = - \beta k^3, \\
\text{eKdV}: &\qquad \omega = - \beta k^3 + \epsilon \beta_1 k^5, \\
 \crr \text{BBM}: &\qquad \crr \omega = - \frac{\beta k^3}{1 + \epsilon \beta k^2},
\end{align}
where $\beta, \beta_1 > 0$ and for simplicity we consider $k \geq 0$. The dispersion relation for the KdV equation is convex, (i.e. there is no inflection point of $\omega = - \beta k^3$ for $k \ge 0$) and hence resonant radiation is not possible. However, the higher order dispersive terms in the eKdV equation introduce non-convexity into the linear dispersion relation and indeed there is an inflection point at the wavenumber  $k = \sqrt{3 \beta / 10 \epsilon \beta_1}$. Hence, the eKdV equation in $(T,\xi)$-variables allows resonant radiation. This is not so for the eKdV equation in slow space variables $(X,\xi)$ where substituting the expression $\eta = \exp i((k + \epsilon \omega)\xi - \omega X)$ {\crr into (\ref{eKdVX})} and truncating the $O(\epsilon^2)$ terms we obtain
\begin{equation}
\omega = - \frac{k^3 (\beta - \epsilon \beta_1 k^2)}{1 + 3 \epsilon \beta k^2}.
\end{equation}
It is then evident for $\beta > 0$, $\beta_1 < 0$ and $-10\beta_1 \geq 3\beta^2$ that there is no inflection point since 
\begin{equation}
\frac{\mathrm{d}^2\omega}{\mathrm{d}k^2} = - \frac{2k}{(1 + 3\epsilon\beta k^2)^3} \left( 3\beta - \epsilon(10 \beta_1 + 3\beta^2) k^2 - 27 \epsilon^2 \beta \beta_1 k^4 ( 1 + \epsilon \beta k^2) \right) \leq 0.
\end{equation}
Hence, the dispersion relation of the slow space eKdV equation for surface water waves is convex, differing from its slow time counterpart. It may therefore be more appropriate to solve the slow space variation where possible.

\begin{figure}
	\centerline{\includegraphics[width=0.955\linewidth]{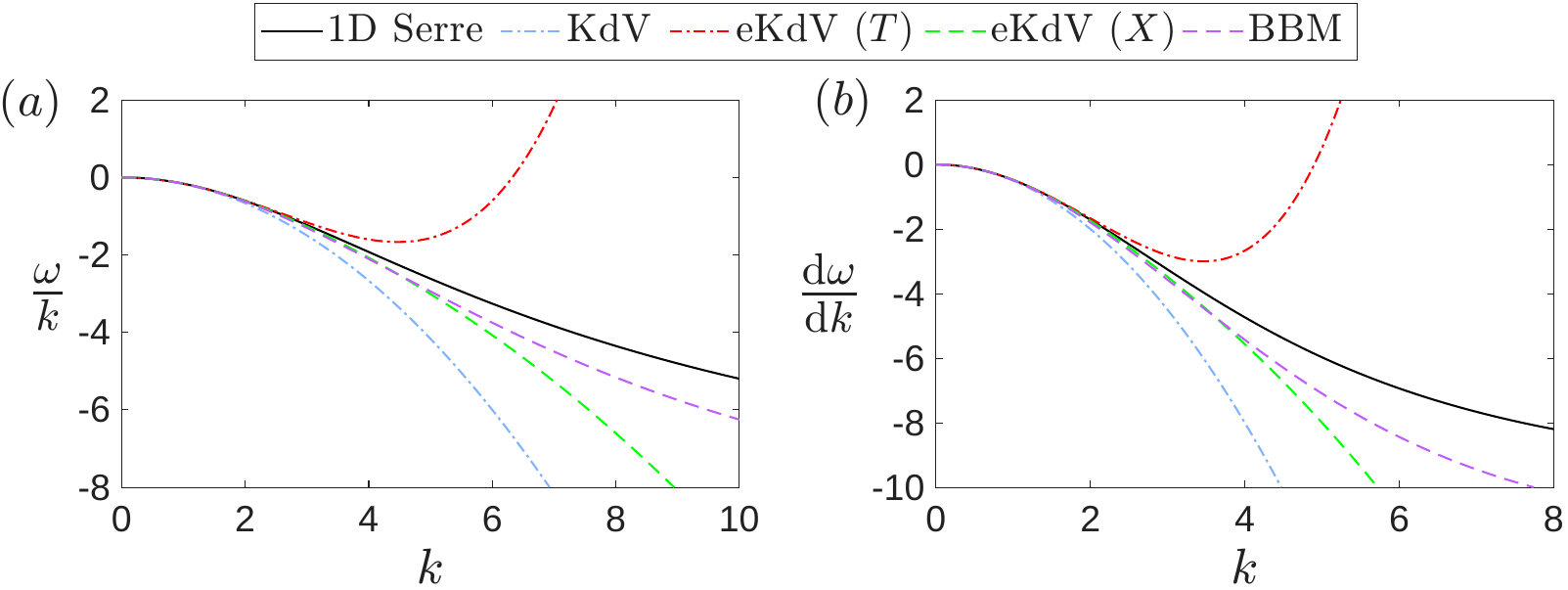}}
	\caption{The phase velocity (\textit{a}) and group velocity (\textit{b}) for $\epsilon = 0.1$ of the KdV equation (blue), eKdV equation in slow time $T$ (red), eKdV equation in slow space $X$ (green), {\crr BBM equation (purple),} and the parent 1D {\crr Serre} equations (black).}
	\label{dispersionplot}
\end{figure}

To determine the linear dispersion relation of the {\crr Serre} equations we linearise (\ref{sgntxi1}) and (\ref{sgntxi2}) and substitute $\eta = A e^{i(k\xi - \omega T)}$ and $u = B e^{i(k\xi - \omega T)}$, {\crr which yields}
the linear dispersion relation
\begin{equation}
\omega = \frac{k}{\epsilon} \left( -1 \pm \sqrt{\frac{3}{3 + \epsilon k^2}} \right), \label{parent_omega}
\end{equation}
where the upper sign describes right propagating waves and the lower sign describes fast left propagating waves in the moving coordinate frame. Physically we are interested in right-propagating waves in the context of the reduced amplitude models, however, the parent systems will also radiate some fast left-propagating radiation. This radiation quickly exits the reference frame and is not of importance here. Resonant radiation is also not possible in the 1D {\crr Serre} equations since for the dispersion curve given by (\ref{parent_omega}) then
\begin{equation}
\frac{\mathrm{d}^2\omega}{\mathrm{d}k^2} = - 9\sqrt{3}k \left( \frac{1}{3 + \epsilon k^2} \right)^{\frac 52} \leq 0.
\end{equation}
We plot both the phase velocity, $v_p = \omega / k$, and the group velocity, $v_g = \mathrm{d}\omega / \mathrm{d}k$, for the KdV, eKdV, {\crr and BBM equations}, and the positive {\crr branch} of (\ref{parent_omega}) in figure \ref{dispersionplot}.

The phase and group velocity for both eKdV formulations match the parent system for larger $k$ than the KdV equation, as expected, however, the slow time formulation clearly gives a non-convex graph. The phase and group velocity of the slow space eKdV equation matches the parent system for larger $k$ than the slow time variation and hence we expect this model to describe dispersive wavetrains more accurately. {\crr The linear dispersion relation of the BBM equation approximates that of the parent 1D Serre equations best, but, of course, it is still rather far from it at large wavenumbers.} In all 4 cases plotted the dispersion relations are unbounded for large $k$ i.e. $\vert \omega(k) \vert \rightarrow \infty$ as $k \rightarrow \infty$, however, {\crr  the phase velocity of the BBM and 1D Serre equations are both bounded by $v_p = - \epsilon^{-1}$. In the numerical scheme used here and given in appendix \ref{appendixB} high wave numbers decay exponentially to machine precision for suitably smooth data and hence we do not observe any noticeable short waves related problems in the KdV equation.}

\begin{figure}
	\centerline{\includegraphics[width=0.95\linewidth]{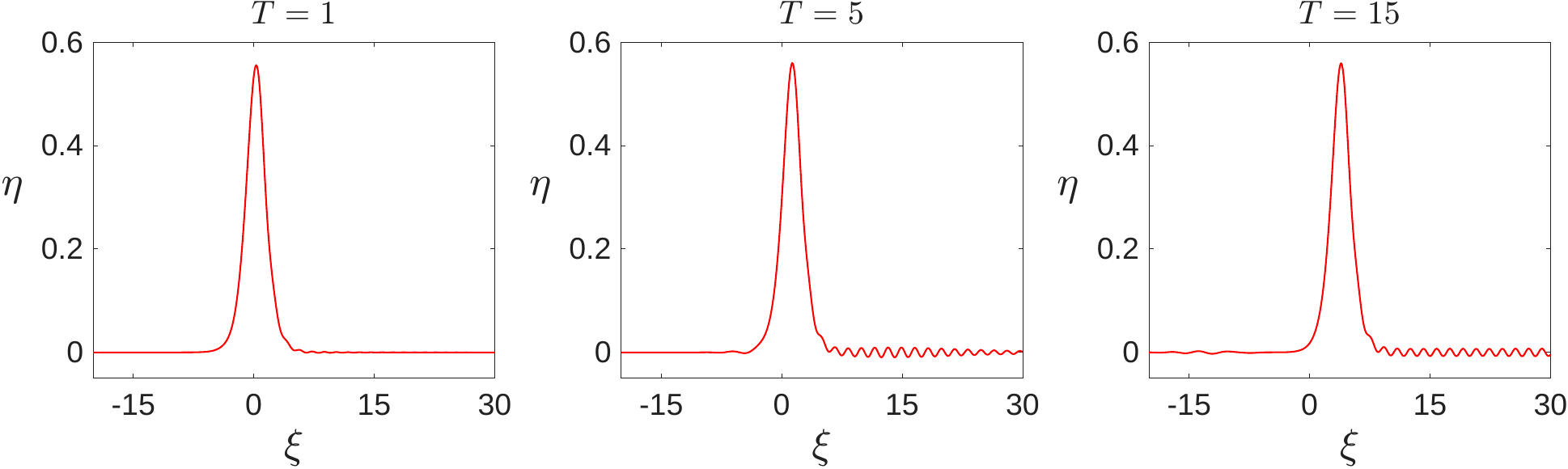}}
	\caption{\normalsize Numerical solution of the eKdV equation for $\epsilon = 0.3$ where the solution is plotted at $T = 1$, $T = 5$, and $T = 15$.}
	\label{resonanceexample}
\end{figure}

To illustrate the resonant radiation produced in the slow time formulation we numerically solve the eKdV equation using the method outlined in appendix \ref{appendixB} for the initial condition 
\begin{equation}
\eta = \operatorname{sech}^2 \left( \frac{\sqrt{3}}{2} ~ \xi \right),
\end{equation}
where $\xi \in [-50,50]$ and $T \in [0,15]$, for $\epsilon = 0.3$, which yields figure \ref{resonanceexample}.

The initial condition is not an exact solution of the eKdV equation and hence some radiation is seen left propagating in the moving coordinate frame. The main feature of interest is the modulated resonant wavetrain in front of the solitary wave. At the early moments of time this feature is seen extending quickly to the right, followed by the larger amplitude solitary wave, as predicted by the shape of the phase and group velocities shown in figure \ref{dispersionplot}. Resonance is always present in the slow time formulation but as $\epsilon$ decreases the amplitude decreases exponentially, see \citet{BGK93} for an analytical description.

The linear dispersion relation of the slow time eKdV equation is hence not appropriate for accurately describing surface water wave behaviour in the entire $\xi$-domain. However, the studies by \citet{W67,FW78} {\crr suggested} that it may be {\crr useful} to keep the nonlinear terms of an equation and replace the dispersive terms with those of the parent system. Whitham {\crr argued} that the KdV equation of the form $\eta_T + \mathcal{L}[ \eta ] + \mathcal{N}[ \eta ] = 0$ can be rewritten as an integro-differential equation
\begin{equation}
\eta_T + \int_{-\infty}^{\infty} K(\xi - \zeta) \eta(T,\zeta) ~ \mathrm{d}\zeta + \mathcal{N}[\eta] = 0, \label{Whithameq1}
\end{equation}
where $\mathcal{L}[\eta]$ and $\mathcal{N}[\eta]$ are the linear and nonlinear terms, respectively, $\zeta$ is a convolution variable, and $K(\xi)$ is a kernel representing the Fourier transform of the linear dispersion relation, $\omega(k)$, where
\begin{equation}
K(\xi) = \frac{1}{2\pi} \int_{-\infty}^{\infty} \omega(k) e^{ik\xi} ~ \mathrm{d}k \label{Kxi}
\end{equation}
(see \citet{W74,LY96}, {\crr also  \citet{KLPS18} for rigorous results}). If we rewrite (\ref{Whithameq1}) in Fourier space then we obtain 
\begin{equation}
\hat \eta_T + i\omega(k) {\hat \eta} +  \mathcal{ \hat N}[\eta] = 0, \label{reducedeta}
\end{equation}
which matches the form of all reduced amplitude models discussed in this study. It is then straightforward to prescribe the linear dispersion relation $\omega(k)$ of the parent system yet preserve the nonlinear terms of the reduced order model. This method has been extensively studied for improving the dispersion properties of the KdV equation, see for example \citet{LY96,MKD15,C18,CHP24}, and the references therein. This equation is known as the KdV--Whitham (KdVW) approximation.

{\crr Our primary aim is to show that that the eKdV equation can be used to describe waves of moderate amplitude.} To regularise the eKdV equation, {\crr and to provide a better description of dispersive radiation},  we take (\ref{eKdV}) in the form of (\ref{Whithameq1}) and replace $\omega(k)$ with (\ref{parent_omega}), the linear dispersion relation of the parent systems. This new equation takes the form
\begin{equation}
\eta_T + \alpha \eta \eta_{\xi} + \epsilon \left( \alpha_1 \eta^2 \eta_{\xi} + \gamma_1 \eta  \eta_{\xi\xi\xi} + \gamma_2 \eta_{\xi} \eta_{\xi\xi} \right) +  \int_{-\infty}^{\infty} K(\xi - \zeta) \eta(T,\zeta) ~ \mathrm{d}\zeta = 0, \label{whithameq} 
\end{equation}
where $K(\xi)$ is the kernel given by (\ref{Kxi}) and the coefficients are unchanged. We call this formulation the extended KdV--Whitham (eKdVW) approximation. To obtain the KdVW equation from (\ref{whithameq}) one should simply set $\alpha_1 = \gamma_1 = \gamma_2 = 0$. For scenarios where the wavelength is short we expect the eKdVW equation to improve the phase error {\crr compared to} the original eKdV equation and for the eKdVW equation to provide a crucial improvement on amplitude and phase of  {\crr solitons of moderate amplitude compared to} the KdVW equation when describing {\crr the evolution of  localised initial data.} This is highlighted in the numerical section that follows.

\section{Derivation of exact and asymptotic solitary wave solutions}
\label{section_solitary_sols}

We have so far derived the eKdV equation and reasoned two different regularisations, we now turn to solutions of the eKdV equation. Unlike the KdV and KdVW equations the eKdV equation is not integrable, except for some special coefficients not considered in this study of generic surface water waves. The eKdV equation can be used to asymptotically approximate soliton solutions of the fully nonlinear and dispersive parent system in various physical contexts, which we test here using the water waves context.

{\crr The KdV equation (\ref{KdV}) and both Gardner equations (\ref{tru_Gardner}) and (\ref{imp_Gardner}) have exact soliton solutions, which are given below, however the KdVW, extended KdV, and extended KdVW equations do not have known exact soliton solutions.  It is known that the KdVW equation does admit periodic travelling waves, see \citet{EK09}. We therefore} produce an asymptotic approximation of a solitary wave solution to the eKdV equation using a near-identity transformation (NIT)  of the eKdV equation to the KdV equation \citep{K85a,K85b,FL96}. {\crr A near-identity transformation is a mapping containing $O(1)$ and $O(\epsilon)$ terms, which in the limit as $\epsilon \rightarrow 0$ reduces to  identity.} Such solutions have been sought by using different NITs to the next member of the KdV hierarchy and the KdV equation by \citet{MS96,MS06, M99} and to the improved Gardner equation by \citep{GBK20, STCK25}. The Gardner equation is formally valid asymptotically when the leading order nonlinear term of the KdV equation is small, and it becomes natural to include the next nonlinearity term in the asymptotic expansion, which is true for internal water waves. This is not the case for surface water waves where the nonlinear coefficient $\alpha = 3/2$ is $O(1)$ and hence it may be more natural to determine a NIT to the KdV equation. This is tested in our numerical runs.

To compare different wave solutions of all the models discussed we construct solutions in $(T,\xi)$-variables and parametrise by the wave speed. The solitary wave solution of the 1D {\crr Serre} equations (\ref{sgntxi1}) and (\ref{sgntxi2}) is determined by scaling the well known solitary wave solution of (\ref{SGN1D1}) and (\ref{SGN1D2}) given by, for example, \citet{DCMM13,MDC17}. The solitary wave solution of (\ref{sgntxi1}) and (\ref{sgntxi2}) has the form
\begin{equation}
\eta(T,\xi) = \frac{c^2 - 1}{\epsilon} \operatorname{sech}^2 \left( \sqrt{\frac{3(c^2 - 1)}{4\delta^2c^2}} (\xi - V T) \right), \qquad u(T,\xi) = \frac{c \eta(T,\xi)}{1 + \epsilon \eta(T,\xi)}, \label{SGNsol}
\end{equation}
where $c = 1 + \epsilon V$ and $V$ is a free parameter which determines the wave speed.

To construct an asymptotic solitary wave solution, accurate to $O(\epsilon^2)$, of the eKdV equation (\ref{eKdV}) we use a Kodama-Fokas-Liu NIT that maps the eKdV equation (\ref{eKdV}) to the KdV equation (\ref{KdV}). For $\eta(T,\xi)$ that satisfies the eKdV equation (\ref{eKdV}), by the NIT written as
\begin{equation}
\eta = B - \epsilon \left[ aB^2 + bB_{\xi\xi} + cB_{\xi} \left( \int_{\xi_0}^{\xi} B(T,\xi') ~ \mathrm{d}\xi' + f(T) \right) + d \xi B_T \right], \label{NIT}
\end{equation}
we obtain the KdV-type equation satisfied by $B(T,\xi)$ as
\begin{equation}
B_T + \alpha BB_{\xi} + \beta B_{\xi\xi\xi} - \epsilon c f'(T) B_{\xi} = 0, \label{KdVf}
\end{equation}
plus the remainder term
\begin{equation}
\epsilon c \left( \frac{\alpha}{2}B(T,\xi_0)^2 + \beta B_{\xi\xi}(T,\xi_0) \right)B_{\xi},
\end{equation}
which we require to be zero. For solitary waves taking sufficient decay conditions as $\xi_0 \rightarrow -\infty$ this is satisfied. We also determine that the NIT coefficients $(a,b,c,d)$ take the values
\begin{align}
&a = \frac{18\beta^2\alpha_1 - 2\alpha^2\beta_1 - 3\alpha\beta\gamma_1}{18\alpha\beta^2}, 
\quad b = \frac{6\beta^2\alpha_1 + \alpha^2\beta_1 - \alpha\beta\gamma_2}{2\alpha^2\beta},  \quad c = \frac{3\beta\gamma_1 - 4\alpha\beta_1}{9\beta^2}, 
\quad d = -\frac{\beta_1}{3\beta^2},
\end{align}
and the function $f(T)$ is a phase shift that is to be determined. We can however remove the function $f'(T)$ from the amplitude equation by virtue of the transformation $\hat\xi = \xi + \epsilon c f(T)$ which means (\ref{KdVf}) can be reduced to the KdV equation
\begin{equation}
B_T + \alpha BB_{\hat\xi} + \beta B_{\hat\xi\hat\xi\hat\xi} = 0. \label{KdVB}
\end{equation}
The KdV equation (\ref{KdVB}) has a one-parameter family of soliton solutions given by
\begin{equation}
B = \frac{3v}{\alpha} \operatorname{sech}^2 \left( \frac 12 \sqrt{\frac{v}{\beta}} \hat\theta \right), \label{KdVsolution}
\end{equation}
where $v$ is a free parameter and $\hat \theta = \hat\xi - vT$, which implies that (\ref{KdVf}) has the solitary wave solution given by (\ref{KdVsolution}) when $\hat\theta = \xi - v T + \epsilon c f(T)$. Applying the NIT given by (\ref{NIT}) to the solitary wave solution of (\ref{KdVf}) yields the approximate eKdV equation solution
\begin{equation}
\eta = M \operatorname{sech}^2 (G\theta) \left[1 - \epsilon F_1 \operatorname{sech}^2 (G\theta) \left(F_2 + F_3 \sinh(2G\theta) + F_4 \cosh(2G\theta) \right)\right], \label{KdVesol}
\end{equation}
where $\theta = \xi - vT + \epsilon c f(T)$ and
\begin{align}
&M = \frac{3v}{\alpha}, \qquad G = \frac 12 \sqrt{\frac{v}{\beta}}, \qquad F_1 = \frac{v}{12\alpha^3\beta^2}, \qquad F_2 = 2\alpha^2 (3\beta(\gamma_1 + \gamma_2) - 13\alpha\beta_1), \\
&F_3 = -\frac{4G\alpha^3}{v} \left( 3\beta^2 c f(T) + v\beta_1\xi \right), \qquad F_4 = \alpha \left( 18\beta^2 \alpha_1 + 19\alpha^2\beta_1 - 3\alpha\beta(4\gamma_1 + \gamma_2) \right).
\end{align}
However, for (\ref{KdVesol}) to be a true travelling wave solution of the eKdV equation it must only be a function of $\theta$ and hence $f(T)$ should be a function such that
\begin{equation}
3\beta^2cf(T) + v\beta_1\xi = C\theta + D,
\end{equation}
from which letting $C = v\beta_1$ and $D = 0$ we obtain
\begin{equation}
3\beta^2 c f(T) = - v^2 \beta_1 T + \epsilon c v \beta_1 f(T),
\end{equation}
and hence
\begin{equation}
f(T) = - \frac{v^2\beta_1}{c(3\beta^2 - \epsilon v \beta_1)}T.
\end{equation}
This function for $f(T)$ is linear in $T$, giving a linear phase shift and simplifying $F_3$ such that $F_3 = -4G\alpha^3 \beta_1 \theta$. The phase shift given by $f(T)$ is an $O(\epsilon)$ change in the speed of the solitary wave solution (\ref{KdVesol}) and so we can reparametrise the solution for a new velocity $V$ such that $\theta = \xi - VT$. To do so we define 
\begin{equation}
v + \frac{\epsilon v^2 \beta_1}{3\beta^2 - \epsilon v \beta_1} = V \quad \Rightarrow \quad v = \frac{3V\beta^2}{3\beta^2 + \epsilon V \beta_1},
\end{equation}
and the second order accurate solitary wave solution to the eKdV equation (\ref{eKdV}) is still given by (\ref{KdVesol}) but now
\begin{equation}
M = \frac{9 V \beta^2}{\alpha ( 3\beta^2 + \epsilon V \beta_1)}, \qquad G = \frac 12 \sqrt{\frac{3 V \beta}{3\beta^2 + \epsilon V \beta_1}}, \qquad F_1 = \frac{V}{4\alpha^3 (3 \beta^2 + \epsilon V \beta_1)}. \label{KdVesolterms}
\end{equation}
The solution given by (\ref{KdVesol}) and (\ref{KdVesolterms}) is now parameterised in the same way as the 1D {\crr Serre} equations solution given by (\ref{SGNsol}).

The approximate solution to the eKdV equation (\ref{eKdV}) obtained from the NIT of the solution to the Gardner equation has been derived by \citet{STCK25}. Here we recap their derivation. Applying the NIT transform (\ref{NIT}) it is possible to obtain the improved Gardner equation
\begin{equation}
B_T + \alpha BB_{\xi} + \beta B_{\xi\xi\xi} + \epsilon \alpha_2 B^2B_{\xi} - \epsilon c f'(T) B_{\xi} = 0, \label{Gardnerf}
\end{equation}
where $f(T)$ is a new function to be determined and the NIT coefficients $(a,b,c,d, \alpha_2)$ take the values
\begin{align}
&a = 0, \qquad b = \frac{5 \alpha \beta_1 + 3 \beta(\gamma_1 - \gamma_2)}{6\alpha \beta}, \qquad c = \frac{3\beta \gamma_1 - 4 \alpha \beta_1}{9\beta^2}, \\
&d = -\frac{\beta_1}{3\beta^2}, \qquad \alpha_2 = \frac{18 \alpha_1 \beta^2 - 2 \beta_1 \alpha^2 - 3\beta \gamma_1 \alpha}{18\beta^2},
\end{align}
plus a remainder term that is also zero provided $\xi_0 \rightarrow -\infty$. The improved Gardner equation (\ref{Gardnerf}) has the solitary wave solution
\begin{equation}
B = \frac{\hat M}{1 + \hat F \cosh(\hat G \theta)}, \label{Gardnersol}
\end{equation}
where $v$ is a free parameter, $\theta = \xi - vT + \epsilon c f(T)$, and
\begin{equation}
\hat F = \sqrt{1 + \frac{6 \epsilon v \alpha_2}{\alpha^2}}, \qquad \hat G = \sqrt{\frac{v}{\beta}}, \qquad \hat M = \frac{6v}{\alpha}.
\end{equation}
The NIT of (\ref{Gardnersol}), which satisfies (\ref{Gardnerf}), was found by \citet{STCK25} to be
\begin{align}
\eta &= \frac{M}{1 + F \cosh (G\theta)} \left[1 - \frac{\epsilon b FG^2 ( F \cosh (2G\theta) - 2 \cosh (G\theta) - 3F )}{2 (1 + F \cosh (G\theta))^2} \right.  \nonumber \\
& \quad + \frac{\epsilon F \sinh (G\theta)}{1 + F\cosh (G\theta)} \left.\left( \frac{2cM \lambda(\theta)}{\sqrt{1 - F^2}} + \frac{G A\beta_1}{3\beta^2 + \epsilon A\beta_1} \theta \right)\right], \label{Gardneresol}
\end{align}
where now $\theta = \xi - VT$ and
\begin{align}
&F = \sqrt{1 + \frac{18\epsilon \alpha_2 V \beta^2}{\alpha^2(3\beta^2 + \epsilon V \beta_1)}}, \qquad 
G = \sqrt{\frac{3V\beta}{3\beta^2 + \epsilon V \beta_1}}, \qquad M = \frac{18V\beta^2}{\alpha(3\beta^2 + \epsilon V\beta_1)}, \\
&f = - \frac{v^2 \beta_1}{c(3\beta^2 - \epsilon \beta_1 v)}T , \qquad \lambda = \operatorname{arctanh} \left[ \sqrt{\frac{1 - F}{1 + F}} \tanh\left(\frac{G\theta}{2}\right) \right].
\end{align}
We have reparameterised the solution by the velocity $V$ compared to the original form. The NIT solution of the improved Gardner equation given by (\ref{Gardneresol}) has a natural restriction on $F$, given $\alpha_2 < 0$, that 
\begin{equation}
V < - \frac{3\alpha^2\beta^2}{\epsilon(18\beta^2\alpha_2 + \alpha^2 \beta_1)}
\end{equation}
to avoid a complex valued amplitude which could be restrictive for moderate $\epsilon$-values. This asymptotic region is outside the formal region of validity for the Gardner equation, however, it is still interesting to see how well the higher order solution performs in this range.

\begin{figure}
	\centerline{\includegraphics[width=0.95\linewidth]{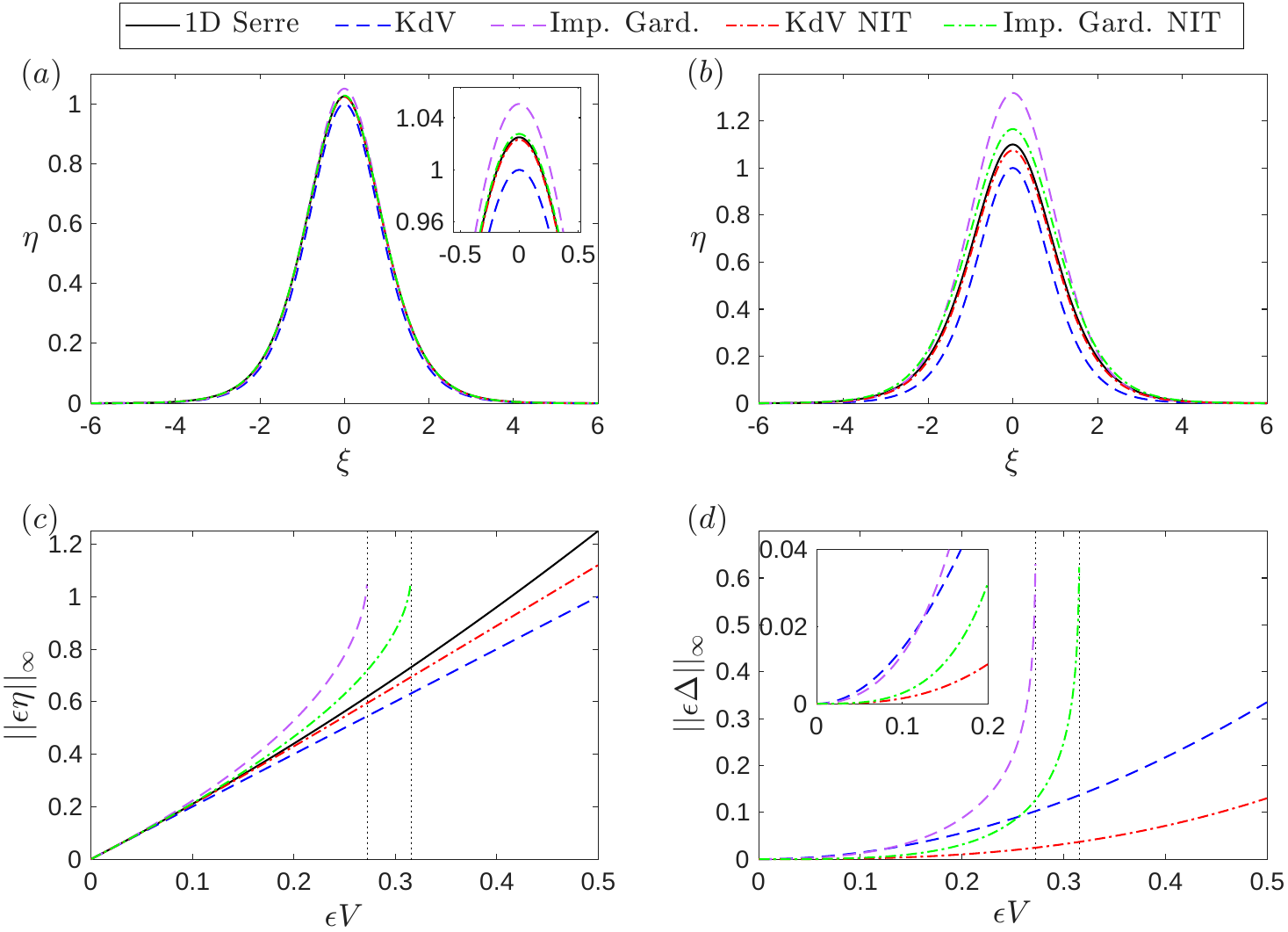}}
	\caption{Analytical plots of the exact 1D {\crr Serre}, KdV, and improved Gardner soliton solutions and the respective NIT solutions are plotted explicitly in ($\textit{a}$) $\epsilon = 0.1$ and ($\textit{b}$) $\epsilon = 0.4$ for $V = 0.5$. In ($\textit{c}$) and ($\textit{d}$) the $L_{\infty}$ norm of $\epsilon \eta$ and $\epsilon \Delta$ is plotted against $\epsilon V$, respectively, where $\Delta$ is the difference between the 1D {\crr Serre} solution and respective reduced order solution.}
	\label{fig_soliton_comps}
\end{figure}

Now that we have the exact 1D {\crr Serre} equations solitary wave solution and two asymptotic solutions to the eKdV equation arising from the NIT of the KdV and improved Gardner equation soliton solutions we compare them analytically in figure \ref{fig_soliton_comps}.

Both the KdV NIT and improved Gardner NIT offer a significant improvement on the KdV and improved Gardner solitary wave solutions. For small $\epsilon V$-values the improved Gardner soliton solution is a better approximation of the 1D {\crr Serre} equation solution than the KdV equation soliton solution but this quickly changes. For all $\epsilon V$-values the KdV NIT solution gives the best approximation, not helped by the improved Gardner equation solutions becoming complex valued at only moderate $\epsilon V$-values. The improved Gardner NIT solution does increase the range of validity to higher $\epsilon$ although it quickly deteriorates below the accuracy of even the KdV equation solution. The KdV NIT solution therefore gives a significant improvement on all three other approximations given here.

\section{Numerical modelling of solitary waves and dispersive radiation}
\label{section_solitary_num}

In this section we compute direct numerical solutions of the 1D {\crr Serre} equations for moderate $\epsilon$-values and compare to the corresponding reduced order models discussed and derived in sections \ref{section_intro} and \ref{section_derivs}. For details on the numerical simulations see appendix \ref{appendixB}. Numerical simulations of localised waves are sought by taking the initial condition for $\eta$ to be NIT of the KdV equation soliton solution given by (\ref{KdVesol}), and the initial velocity for the 1D {\crr Serre} equations is given by (\ref{u}). 

\begin{figure}
	\centerline{\includegraphics[width=0.90\linewidth]{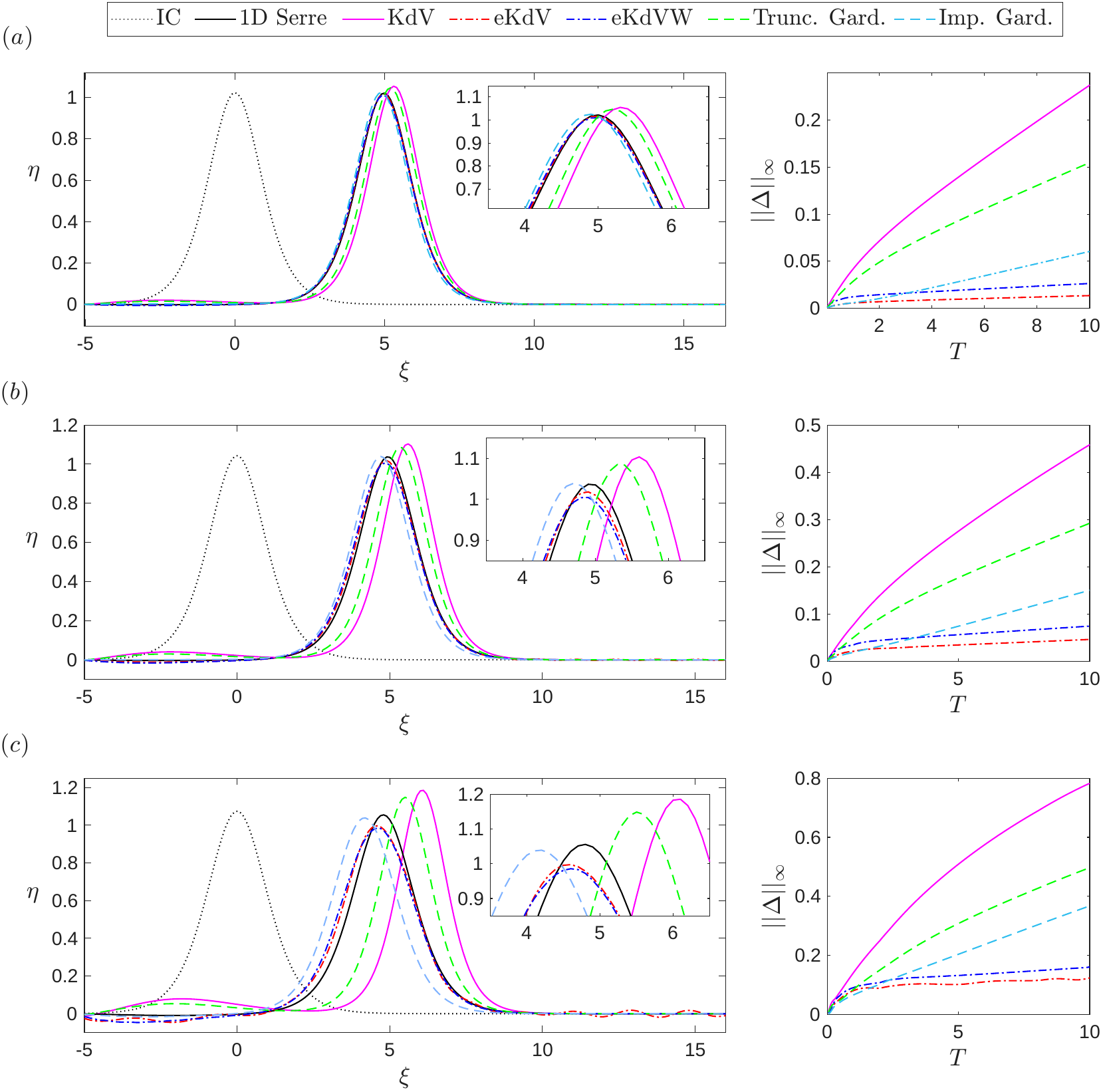}}
	\caption{Numerical solutions of the 1D {\crr Serre} equations and the corresponding reduced amplitude models where the left column gives the amplitude comparison and the right column gives the $L_{\infty}$ norm of the difference, $\Delta$, between the parent model and reduced model. {\crr The initial condition (IC) is plotted at $T = 0$ and} solutions are plotted at $T = 10$ for $V = 0.5$ where (\textit{a}) $\epsilon = 0.1$, (\textit{b}) $\epsilon = 0.2$, and (\textit{c}) $\epsilon = 0.4$.}
	\label{fig_soliton_sgn_1.5}
\end{figure}

First we observe the evolution of localised initial conditions with a positive amplitude. Using (\ref{KdVesol}) with $V = 0.5$ we solve the reduced order models and 1D {\crr Serre} equations where $\xi \in [-50,50]$ and $T \in [0,10]$, for $\epsilon = 0.1, 0.2$, and $0.4$. The results are shown in  figure \ref{fig_soliton_sgn_1.5}.

As can be seen in figure \ref{fig_soliton_sgn_1.5} the eKdV equation produces resonant radiation in front of the solitary wave {\crr which is expected since we do not initiate the simulations with an exact soliton solution of the eKdV equation}. This feature is present in all simulations presented but the amplitude is dependent upon $\epsilon$ and is only noticeable when $\epsilon$ becomes large. Despite this the eKdV equation performs well in the rest of the domain. Interestingly, the resonant radiation produced by the eKdV equation was seen to be smaller when using the coefficients determined by the 1D {\crr Serre} equations and larger when using the coefficients derived from the 1D BP equations. The KdV equation is not expected to work in this large $\epsilon$ range and indeed sees the largest error, however, the shape and amplitude of the KdV solution are qualitatively correct, the main error is a substantial phase shift. Both Gardner equations perform better than the KdV equation but are considerably worse than the eKdV and eKdVW equations. The key difference being  {\crr the solution of} the truncated Gardner equation appears faster and taller than {\crr that of} the 1D {\crr Serre} equations, and {\crr the solution of} the improved Gardner equation has the correct amplitude but appears slower. {\crr Interestingly, despite being formally argued when the nonlinear coefficient is small or vanishing, which is not the case for surface water waves where $\alpha = 3/2$ is $O(1)$, both Gardner equations still offer a significant improvement compared to the KdV equation. The eKdVW and eKdV equations work considerably better than even the improved Gardner equation.} Also, the eKdVW equation proves to be a good regularisation for the eKdV equation in the long time evolution of solitary waves since there is no resonance and it has only a small error in the amplitude and phase even for large $\epsilon$-values when compared to the eKdV equation solution.

\begin{figure}
	\centerline{\includegraphics[width=0.90\linewidth]{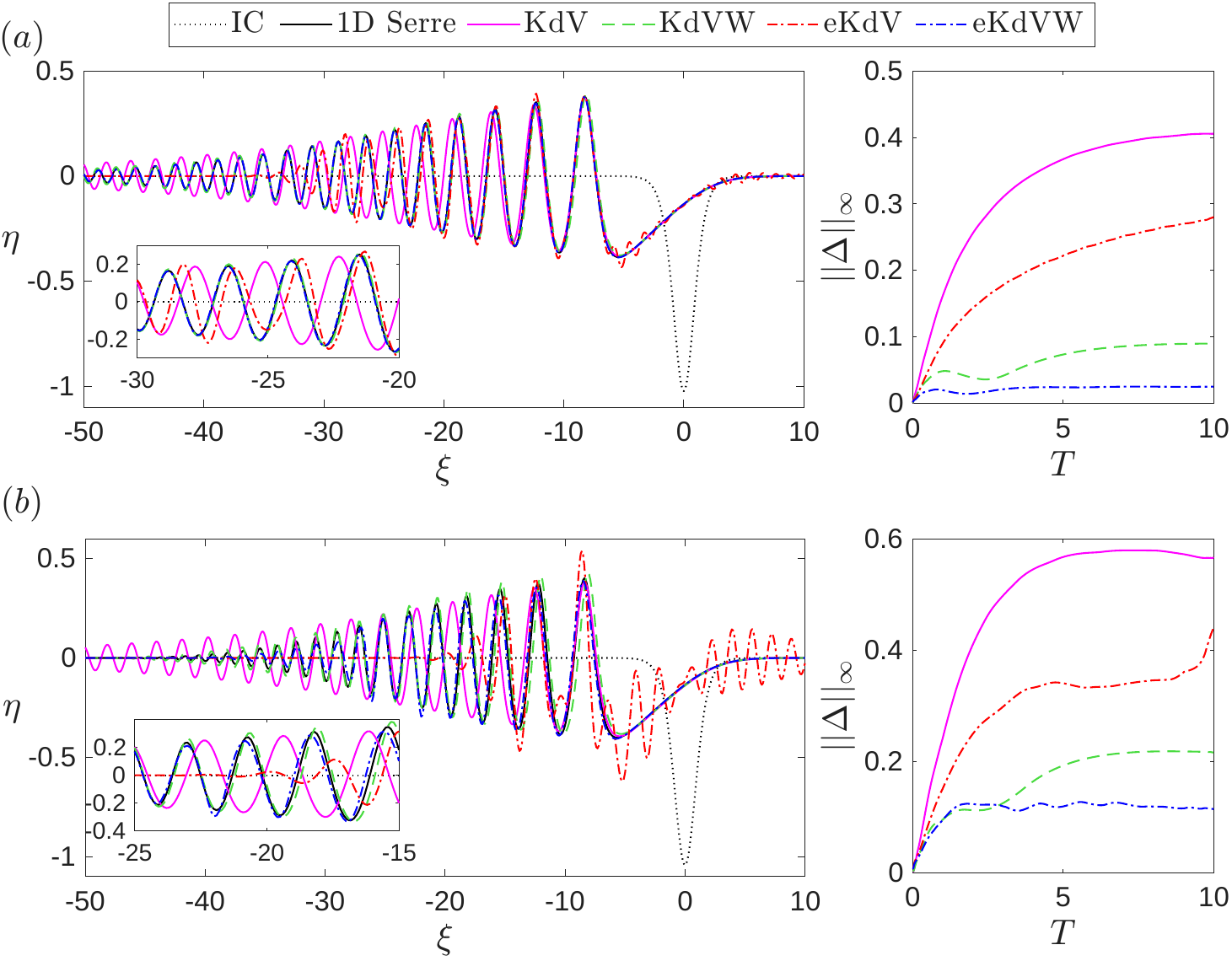}}
	\caption{Numerical solutions the 1D {\crr Serre} equations and the corresponding reduced amplitude models where the left column gives the amplitude comparison and the right column gives the $L_{\infty}$ norm of the difference, $\Delta$, between the parent model and reduced model. {\crr The initial condition (IC) is plotted at $T = 0$ and} solutions are plotted at $T = 10$ for $V = 0.5$ where (\textit{a}) $\epsilon = 0.1$ and (\textit{b}) $\epsilon = 0.2$.}
	\label{fig_negsoliton_sgn_1.5}
\end{figure}

Next, we continue to use (\ref{KdVesol}) for $V = 0.5$, but map $\eta \rightarrow -\eta$ which gives an initial condition of negative amplitude. For surface water waves of negative amplitude no solitary waves {\crr are generated} and the evolution will form a purely dispersive wave train. We solve the reduced order models and 1D {\crr Serre} equations where $\xi \in [-100,50]$ and $T \in [0,10]$, for $\epsilon = 0.1$ and $0.2$. The results are shown in figure \ref{fig_negsoliton_sgn_1.5}. We solve this case for smaller $\epsilon$-values as the stability properties of the eKdV equation are worse for negative amplitudes.

In the initial evolution the localised wave of negative amplitude disintegrates into a dispersive wave train where the wavenumber of the leading edge is still close to zero but increases as we move left in the domain. In the tail there are higher frequency, smaller linear waves. For this case nonlinearity is less important and it is therefore expected that the KdV, and Gardner equations will all perform similarly since all three have the same linear dispersion relation, and hence the same behaviour in this trailing edge. We therefore only plot the KdV equation results. In general for these three models there is a significant phase shift in the tail but the leading edge is described well. {\crr Next, the eKdVW equation performs much better than the eKdV equation across the whole domain in both cases shown here since it has the exact linear dispersion relation.} When $\epsilon = 0.1$ the resonance of the eKdV equation is small and has only a small phase shift in the tail, however, when $\epsilon = 0.2$ the resonance destroys the oscillatory tail and gives very poor agreement in the entire domain. The eKdVW equation performs exceptionally {\crr well} for both values of $\epsilon$ and is indistinguishable from the parent models in both simulations shown in figure \ref{fig_negsoliton_sgn_1.5}. The eKdVW solution is a significantly cheaper solution to compute numerically than 1D {\crr Serre} equations. 

\begin{figure}
	\centerline{\includegraphics[width=0.90\linewidth]{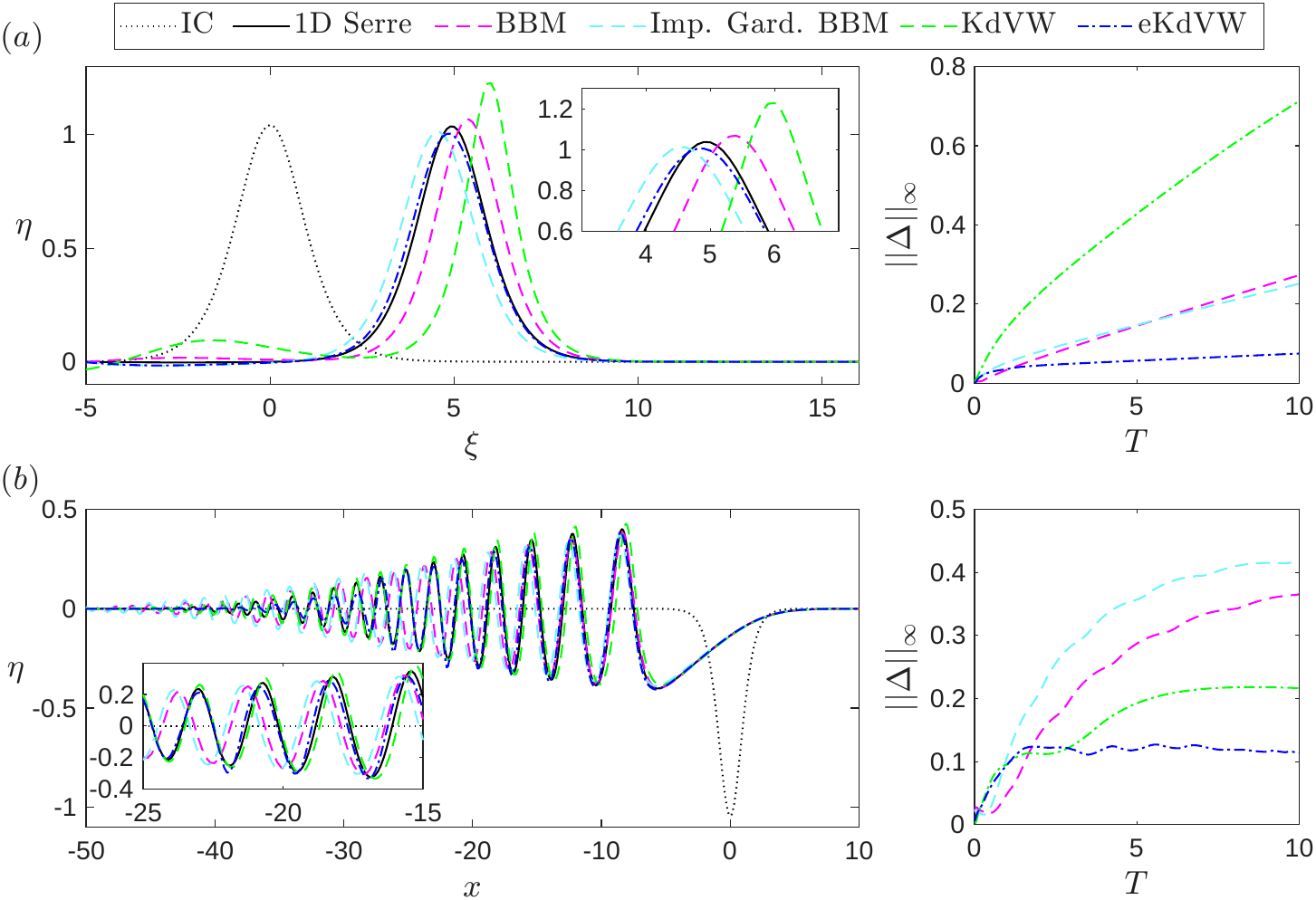}}
	\caption{{\crr Numerical solutions the 1D Serre equations and the corresponding reduced amplitude models where the left column gives the amplitude comparison and the right column gives the $L_{\infty}$ norm of the difference, $\Delta$, between the parent model and reduced model. The initial condition (IC) is plotted at $T = 0$ and solutions are plotted at $T = 10$ for $V = 0.5$ and $\epsilon = 0.2$ where (\textit{a}) the initial condition is $\eta$ and (\textit{b}) $\eta \rightarrow -\eta$.}}
	\label{fig_BBM}
\end{figure}

{\crr Finally, we demonstrate the difference between the regularisations of the BBM equation (\ref{BBM}) and the eKdVW equation (\ref{whithameq}). Given the improvement both Gardner equations, truncated and improved, made on the usual KdV equation in this regime of moderate nonlinearity we also consider the performance of the improved Gardner BBM equation given in the fixed reference frame as
\begin{equation}
 \eta_t + \eta_x + \epsilon ( \alpha \eta\eta_{x} - \beta \eta_{xxt} + \epsilon \alpha_2 \eta^2 \eta_x ) = 0.
\end{equation}
We again initiate simulations using (\ref{KdVesol}) for $V = 0.5$ firstly for a wave with positive amplitude and secondly mapping $\eta \rightarrow -\eta$. We solve all models for positive initial data using $\xi \in [-50,50]$ and for negative initial data using $\xi \in [-100,50]$. We take $T \in [0,10]$ and the BBM models are solved using $t = [0,10/\epsilon]$, for $\epsilon = 0.2$. The results are shown in figure \ref{fig_BBM}.}

{\crr For the initial condition with positive amplitude both BBM  models provide a qualitatively correct description with either a slight positive or negative phase shift. Both of these models are considerably better than the KdVW equation but do not perform as well as the eKdVW equation. For the initial condition with negative amplitude the description of the dispersive wave train is again best described by the eKdVW equation and instead the KdVW equation performs better than the BBM equations. Comparing the errors to figure \ref{fig_soliton_sgn_1.5} and \ref{fig_negsoliton_sgn_1.5}, however, shows that the BBM equations both make a significant improvement on the KdV equation. In these runs there was no significant improvement made by introducing the improved Gardner BBM equation over the usual BBM equation.}

\subsection{Regularisation in slow space}

Here we demonstrate that it is possible to regularise the eKdV equation by casting the problem in $(X,\xi)$-variables where {\crr there is no resonant radiation.} The spatial Whitham equation has been used  by \citet{TKCO16, CHP24}. To cast the problem into these variables numerically we initiate the 1D {\crr Serre} equations and solve for the domains $t \in [t_{\mathrm{min}},t_{\mathrm{max}}]$ and $\xi \in [\xi_{\mathrm{min}},\xi_{\mathrm{max}}]$. Data is simultaneously extracted along the curve $X_{\mathrm{min}} = \epsilon( \xi + t )$ and passed to the reduced order models as an initial condition where solutions are calculated for $X \in [X_{\mathrm{min}},X_{\mathrm{max}}]$ across the same domain for $\xi$. To ensure that data can be extracted correctly the domains for the 1D {\crr Serre} equations should be chosen such that $X \in [\epsilon(\xi_{\mathrm{max}} + t_{\mathrm{min}}), \epsilon(\xi_{\mathrm{min}} + t_{\mathrm{max}})]$. We then make comparisons at $X = X_{\mathrm{max}}$.

\begin{figure}
	\centerline{\includegraphics[width=0.90\linewidth]{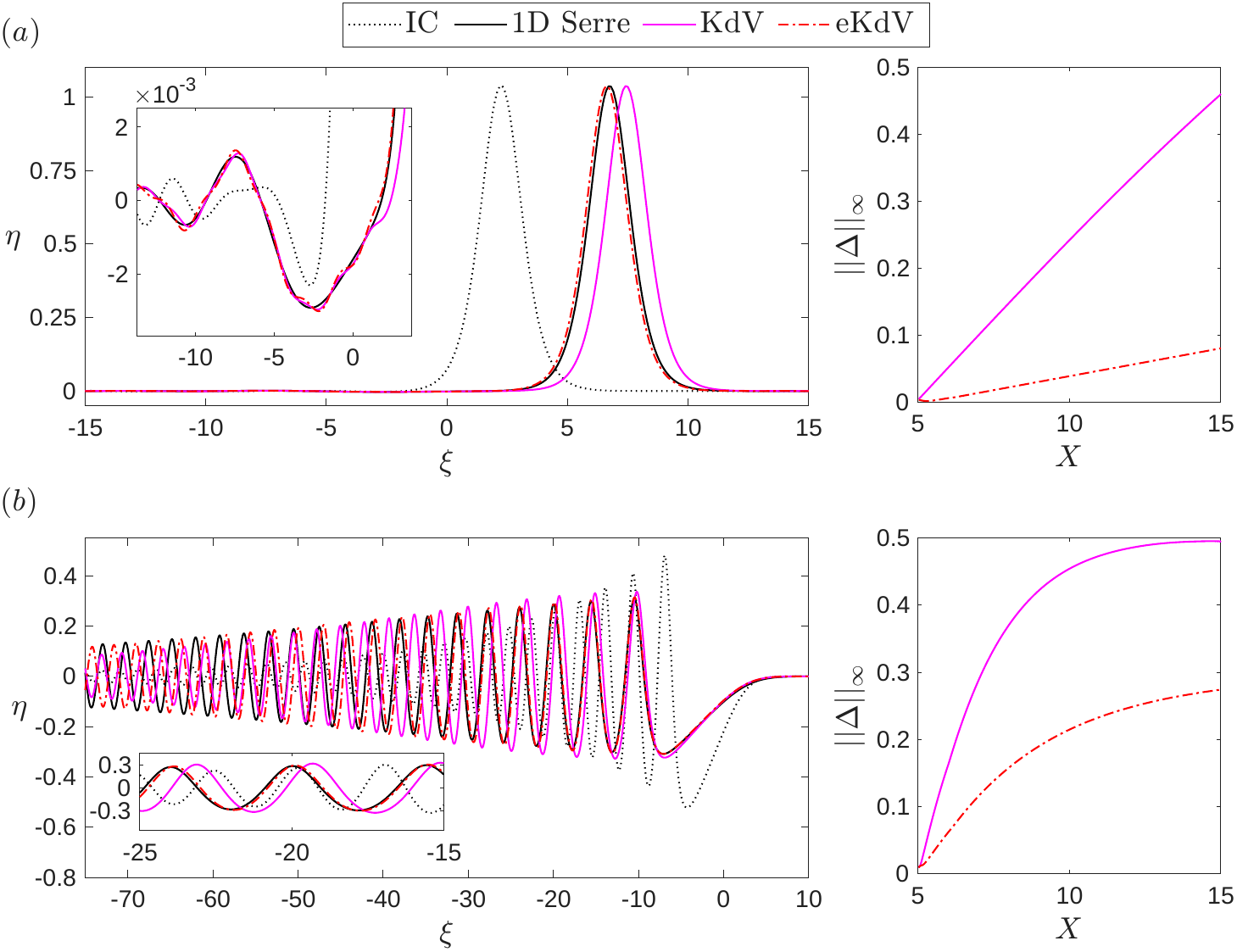}}
	\caption{Numerical solutions of the 1D {\crr Serre} equations and the corresponding reduced amplitude models where the left column gives the amplitude comparison and the right column gives the $L_{\infty}$ norm of the difference, $\Delta$, between the parent model and reduced model. {\crr The initial condition (IC) is plotted at $X = 5$ and} solutions are plotted at $X = 15$ for $V = 0.5$ and $\epsilon = 0.2$ where (\textit{a}) the initial condition is $\eta$ and (\textit{b}) $\eta \rightarrow - \eta$.}
	\label{fig_X_comp}
\end{figure}

Initiating the 1D {\crr Serre} equations with (\ref{KdVesol}) firstly for a wave with positive amplitude and secondly mapping $\eta \rightarrow - \eta$ we obtain figure \ref{fig_X_comp}. The domains for the 1D {\crr Serre} equations are taken to be $t \in [0,175]$ and $\xi \in [-100,25]$ for $\epsilon = 0.2$. Data is extracted along the curve $X_{\mathrm{min}} = \epsilon (\xi + t) = 5$ and passed to the reduced order models where we solve for the domains $X \in [5,15]$ and $\xi \in [-100,25]$. We do not solve for the truncated and improved Gardner equations since they {\crr produce results close to that of} the KdV equation.

For the initial condition with positive amplitude the eKdV equation gives an important correction on the phase to the KdV equation and gives a more accurate description of the small oscillatory tail. The solitary wave of the eKdV equation has a small negative phase shift compared to the 1D {\crr Serre} equations for this large value of $\epsilon$ and very long propagation time but the phase is in considerably better agreement than that of the KdV equation. For the initial condition of negative amplitude the eKdV equation again demonstrates a better description than the KdV equation of the leading edge and dispersive wave train behind where the KdV equation quickly becomes out of phase. For large wavenumbers the eKdV equation begins to fail and this is explained by the increasing discrepancy between the linear dispersion relations of the 1D {\crr Serre} and eKdV equations as the wavenumber increases.

\subsection{Predicting a reduced model}

In the previous sections it has been shown that the best reduced model depends strongly upon the evolution scenario, supporting the conclusions of previous studies using the Whitham equation (e.g. \cite{MKD15}). For initial data where the developing solution contains strong dispersive radiation the eKdVW equation is a better model than the eKdV equation and vice versa for cases when the initial condition evolves mainly into solitary waves.  It would therefore be beneficial to understand key properties of the evolution \textit{a priori} without having to construct and study several different models and solutions. In all cases presented here the KdV equation provides a qualitatively correct description of the evolution. Therefore, by exploiting the known conservation laws, and constructing analytical solutions of the KdV equation via the IST, we can determine the evolution scenario and predict the reduced model that will provide the best approximation of the 1D {\crr Serre} equations.

\begin{figure}
	\centerline{\includegraphics[width=0.90\linewidth]{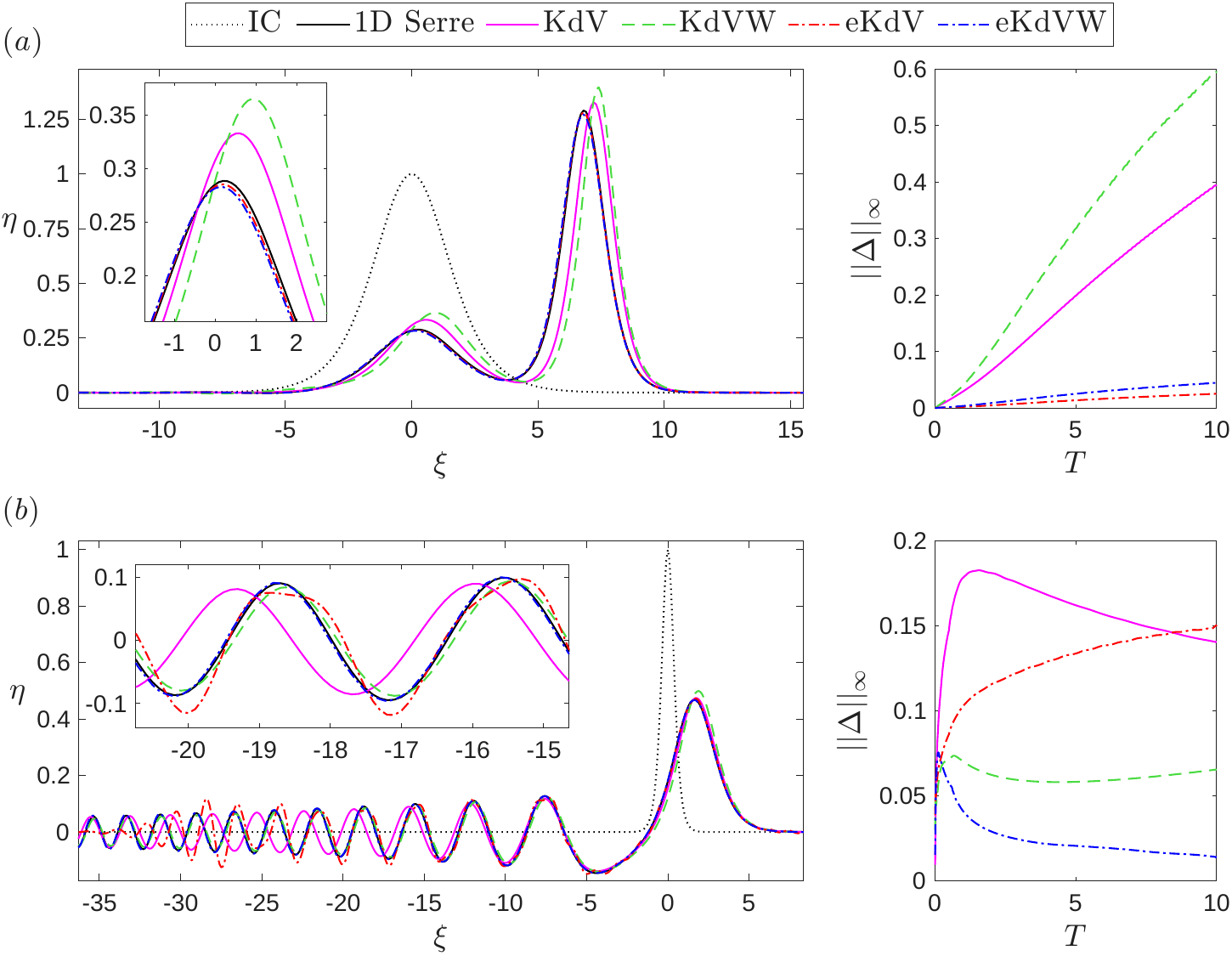}}
	\caption{Numerical solutions of the 1D {\crr Serre} equations and the corresponding reduced amplitude models where the left column gives the amplitude comparison and the right column gives the $L_{\infty}$ norm of the difference, $\Delta$, between the parent model and reduced model. {\crr The initial condition (IC) is plotted at $T = 0$ and} solutions are plotted at $T = 10$ for $\epsilon = 0.1$ where (\textit{a}) $b = 0.5$ and (\textit{b}) $b = 2$.}
	\label{fig_IST_comp}
\end{figure}

The first 3 of an infinite number of conservation laws of the KdV equation (\ref{KdV}) are {\crr given by}
\begin{align}
\frac{\mathrm{d}\mathcal{Q}_1}{\mathrm{d}T} = \frac{\mathrm{d}}{\mathrm{d}T} \int_{-\infty}^{\infty} \eta ~ \mathrm{d}\xi = 0, \quad \frac{\mathrm{d}\mathcal{Q}_2}{\mathrm{d}T} = \frac{\mathrm{d}}{\mathrm{d}T} \int_{-\infty}^{\infty} \eta^2 ~ \mathrm{d}\xi = 0, \quad \frac{\mathrm{d}\mathcal{Q}_3}{\mathrm{d}T} = \frac{\mathrm{d}}{\mathrm{d}T} \int_{-\infty}^{\infty} \left (\eta^3 - \frac 13 \eta_{\xi}^2\right ) ~ \mathrm{d}\xi = 0, \label{conservationl}
\end{align}
 \citep{BK67, MGK68, AS79}. {\crr The quantity $\mathcal{Q}_1$ corresponds to the \textit{excess} mass not including the mass of the base state of the fluid, and the mechanical momentum and energy are linear combinations of $\mathcal{Q}_1$,$\mathcal{Q}_2$, and $\mathcal{Q}_1$,$\mathcal{Q}_2$,$\mathcal{Q}_3$, respectively \citep{AK14}.} We consider the initial condition for the KdV equation given by
\begin{equation}
\eta = a \operatorname{sech}^2 (b\xi), \label{KdV_IC}
\end{equation}
where $a$ and $b$ determine the wave amplitude and wavelength, respectively. To infer the results of the IST we cast the KdV equation (\ref{KdV}) into the canonical form 
\begin{equation}
U_{\tau} - 6UU_{\zeta} + U_{\zeta\zeta\zeta} = 0,
\end{equation}
using the change of variables $\eta = -2U/3$, $T = 6\tau$ and $\xi = \zeta$. The wavefield is defined by the spectrum of the Schr\"{o}dinger equation
\begin{equation}
\Psi_{XX} + \left[\lambda - U(\zeta) \right]\Psi = 0, \label{Schrodinger_eq}
\end{equation}
where the potential is given by the initial condition (\ref{KdV_IC}), which in scaled variables is given as
\begin{equation}
U(\zeta) = - \Lambda \operatorname{sech}^2 \left(\frac{\zeta}{L} \right),
\end{equation}
where $\Lambda = 3a/2$ and $L = 1/b$. The number of solitons generated in the evolution is given by the greatest integer satisfying the inequality
\begin{equation}
N < \frac{1}{2} \left[\left(1 + 4\Lambda L^2 \right)^{\frac{1}{2}} + 1 \right] \label{soliton_n}
\end{equation}
\citep{LL59}. If the initial condition is chosen such that $3a = 2b^2 N(N+1)$ then the reflection coefficient is zero and the evolution is purely solitons with no radiation (e.g. \citet{D89}). The discrete eigenvalues are given by
\begin{equation}
\lambda = -k_n^2,
\end{equation}
where
\begin{equation}
k_n = \frac{1}{2L}\left[\left(1 + 4\Lambda L^2\right)^{\frac{1}{2}} - (2n-1) \right] > 0 \qquad \text{for} \qquad n = 1,2,\ldots,N. \label{ISTk}
\end{equation}
The long-time asymptotics takes the form
\begin{equation}
U\left( \tau,\zeta \right) \simeq - \sum_{n=1}^N 2k_n^2 \operatorname{sech}^2 \left( k_n(\zeta - 4k_n^2\tau - \zeta_n)\right) + \text{radiation}, \label{ISTU}
\end{equation}
where for completeness we give the phase shifts, written as
\begin{equation}
\zeta_1 = \frac{1}{2k_1} \ln \left(\frac{c_1}{2k_1} \right), \qquad
\zeta_n = \frac{1}{2k_n} \ln \left(\frac{c_n}{2k_n} \prod_{m=1}^{n-1} \left( \frac{k_n - k_m}{k_n + k_m} \right)^2 \right), 
\qquad n > 1,
\end{equation}
and the constants $c_n$ are found as
\begin{equation}
c_n = \left(\int_{-\infty}^{\infty} \Psi_n^2(x) \text{ d}x \right)^{-1},
\end{equation}
where $\Psi_n(x)$ is the eigenfunction of (\ref{Schrodinger_eq}) corresponding to the $n^{\text{th}}$ eigenvalue,  $-k_n^2$:
\begin{equation}
\Psi_n = \text{const}\left(1 - \tanh^2 \frac{x}{L} \right)^{\frac{k_nL}{2}} {}_{2}F_{1}\left(1-n, 2k_nL + n,k_nL+1, \frac{1 - \tanh \frac{x}{L}}{2} \right).
\end{equation}
Here, ${}_{2}F_{1}(\cdots)$ is the hypergeometric function, and the constant should be chosen to normalise the eigenfunction at  infinity,
\begin{equation}
\Psi_n \sim \exp^{-k_nx} \quad \text{as} \quad x \rightarrow + \infty.
\end{equation}
In the original problem variables the long time asymptotics is given as
\begin{equation}
\eta\left( T,\xi \right) \simeq \frac 43 \sum_{n=1}^N k_n^2 \operatorname{sech}^2 \left( k_n \left(\xi - \frac 23 k_n^2 T - \zeta_n \right)\right) + \text{radiation}, \label{ISTeta}
\end{equation}
and the conserved quantities in (\ref{conservationl}) can be calculated. In the long time evolution the solitons are suitably separated and ordered by height with only an exponentially small crossover between them.  Therefore, the contribution from the solitonic part can be estimated by a summation of integrals for each $N$ solitons individually and it is not necessary to calculate the phase shifts explicitly. Using (\ref{KdV_IC}) we estimate the contribution of radiation in the {\crr 
three conserved quantities in (\ref{conservationl})} as the difference between the conserved quantity in the initial condition and the contribution from the solitonic part known from the IST solution. This can be given as
\begin{align}
\mathcal{Q}_{1r} = \frac{2a}{b} - \sum_{n=1}^{N} \frac{8k_n}{3}, \qquad \mathcal{Q}_{2r} = \frac{4a^2}{3b} - \sum_{n=1}^{N} \frac{64 k_n^3}{27}, \qquad \mathcal{Q}_{3r} = \frac{16a^2(3a-b^2)}{45b} - \sum_{n=1}^{N} \frac{256 k_n^5}{135},
\end{align}
respectively. We show two examples of this for $a = 1$ and $b = 1/2,2$. When $b = 1/2$ the initial condition evolves in the KdV regime into two solitons, as indicated by (\ref{soliton_n}), and (\ref{ISTk}) yields that $k_1 = 1$ and $k_2 = 1/2$. Interestingly for this initial condition $a,b$ satisfy the relation $3a = 2b^2 N(N+1)$ for $N=2$ and hence $\mathcal{Q}_{1r} = \mathcal{Q}_{2r} = \mathcal{Q}_{3r} = 0$. Since there is no radiation in the KdV equation we expect the eKdV equation to perform the best. When $b = 2$ there is only one soliton present with the eigenvalue determined by $k_1 = \sqrt{5/2} - 1$. For this case, denoting quantities of the initial condition with subscript `I' means we determine that $\mathcal{Q}_{1r} \sim -0.55 \mathcal{Q}_{1I}$, $\mathcal{Q}_{2r} \sim 0.30 \mathcal{Q}_{2I}$, and $\mathcal{Q}_{3r} \sim 1.70 \mathcal{Q}_{3I}$. Therefore, a significant {\crr portion} of the {\crr conserved quantities} are {\crr contained} in the radiation and we expect the KdVW and eKdVW to perform better than the eKdV equation. To illustrate this we show the numerical evolution of the initial condition (\ref{KdV_IC}) in figure \ref{fig_IST_comp} taking the parameters $a = 1$, $b = 2,1/2$, and $\epsilon = 0.1$, where the domains are taken to be $T \in [0,10]$ and $\xi \in [-100,50]$.

When $b = 1/2$ the initial condition evolves into a two soliton solution of the KdV equation (\ref{KdV}), and hence there is no radiation since all three of the {\crr 
conserved quantities} are {\crr contained}  in the solitons. The IST solution amplitudes, $4k_n^2/3$, are a slight overestimate of the 1D {\crr Serre} equation evolution, although qualitatively correct. For the small $\epsilon$-value here the resonance in the eKdV equation is small and the approximation given by the eKdV equation is the most accurate and as expected the eKdVW equation performs second best. However, when $b = 2$ the IST indicates that a significant {\crr portion} of the {\crr conserved quantities} are {\crr contained} in the radiation and indeed the eKdVW is the best model. In the evolution the eKdVW equation is almost exactly replicating the 1D {\crr Serre} equations with the KdVW being a close second. Both the KdV and eKdV equations perform poorly in the tail but offer a qualitatively correct description of the {\crr lead} soliton. For this case the initial evolution of the 1D {\crr Serre} equations is highly dispersive and the parent model evolves differently to the reduced models which gives the initial jump in the $L_{\infty}$ norm for all reduced models.

\section{Conclusions}
\label{section_conclusion}

In this study we investigated the evolution of moderately nonlinear surface water waves in the scope of the KdV, eKdV, and Gardner equations compared to the 1D {\crr Serre} equations \citep{S53,SG69, GN76}. We introduced the extended KdV--Whitham approximation as an extension of the KdV--Whitham approximation \citep{W67,FW78,W74} to regularise the eKdV equation and determined, using a NIT transformation \citep{K85a,K85b,FL96}, an asymptotic solution of the eKdV equation approximating the solitary wave solution of the 1D {\crr Serre} equations accurately. {\crr We also compared the eKdVW equation to the BBM equation and the improved Gardner BBM equation to highlight the benefit of modelling with the eKdVW equation.}

For the coefficients of surface water waves the eKdV equation in the slow time variable, $T = \epsilon t$, exhibits resonant radiation that we showed is not present in either of the parent 1D {\crr Serre} equations or the slow space, $X = \epsilon x$, formulation of the eKdV equation. In the interest of modelling long nonlinear waves it is important to regularise the eKdV equation and we have shown that this is possible in two ways. The first is to cast the problem into the slow space formulation where for moderate $\epsilon$ values we showed the description of  the evolution of localised positive and negative amplitude initial conditions is greatly improved compared to the KdV equation. Secondly, if the slow time evolution variable is kept then we showed the extended KdV--Whitham equation, given by
\begin{equation}
\eta_T + \alpha \eta \eta_{\xi} + \epsilon \left( \alpha_1 \eta^2 \eta_{\xi} + \gamma_1 \eta  \eta_{\xi\xi\xi} + \gamma_2 \eta_{\xi} \eta_{\xi\xi} \right) +  \int_{-\infty}^{\infty} K(\xi - \zeta) \eta(T,\zeta) ~ \mathrm{d}\zeta = 0,
\end{equation}
regularises the eKdV equation well for positive localised initial conditions with only a minor error in phase and amplitude of the crest compared to the eKdV equation. However, for negative localised initial conditions the eKdVW equation significantly outperforms the eKdV equation and {\crr the evolution} is largely indistinguishable from {\crr that in} the parent system.  In all cases tested the eKdVW equation was either first or second in accuracy and even the first order KdVW equation was preferable for large $\epsilon$ simulations of negative initial data. {\crr The BBM equation was better than the KdV equation for both positive and negative initial data of moderate amplitude, but significantly worse that the eKdVW equation.}

Due to the most accurate reduced order model being directly linked to the amount of radiation seen in the evolution we showed that it can be useful to determine the evolution scenario \textit{a priori} by employing the IST \citep{GGKM67,GGKM74} and the {\crr first three} conserved quantities of the KdV equation \citep{BK67, MGK68}. It is then necessary to determine the difference in conserved quantities between the initial condition and the long time asymptotics in turn to {\crr estimate} the {\crr portion of the conserved quantity} present in the radiation. When the quantities of the radiation are significant it is best to solve the eKdVW equation. 

Replacing the linear dispersion relation with a more preferable one greatly improved the description of the problems tested here. {\crr We also note that although the eKdVW equation is not the best model for the description of  solitary waves of moderate amplitude, the difference between the eKdV description (best) and eKdVW description (second best) is small } (see also a discussion in the leading order internal wave case by \citet{LY96}). {\crr We note that the present studies  may find useful applications and extensions including even higher nonlinearities} due to universality of the mathematical models (see, for example, \citet{HFS22}, {\crr \citet{NK24}, \citet{Melnikov}}, and the references therein). 
\bigskip

\noindent
{\bf \large Acknowledgments}
\bigskip

\noindent
The authors would like to thank the referees for their helpful comments. Karima Khusnutdinova is grateful to the organisers of the conference “Wave Dynamics, Integrability and
Beyond” (WaDIB 2025) for the invitation to participate and stimulating discussions during the conference.
\bigskip

\noindent
{\bf \large Conflicts of Interest}
\bigskip

\noindent
The authors declare no conflicts of interest.
\bigskip

\noindent
{\bf \large Data Availability Statement}
\bigskip

\noindent
No datasets were generated or analysed during the current study.

\numberwithin{equation}{section}
\appendix
\section{Derivation of the eKdV equation from the 1D BP equations}
\label{appendixA}

The eKdV equation can similarly be derived using asymptotic multiple scales from the 1D {\crr Boussinesq-Peregrine (BP)} equations (\ref{Bouss1D1}) and (\ref{Bouss1D2}). Applying the change of variables $(t,x) \rightarrow (T,\xi)$, where $T = \epsilon t$ and $\xi = x-t$ to (\ref{Bouss1D1}) and (\ref{Bouss1D2}) yields 
\begin{align}
&u_{\xi} - \eta_{\xi} + \epsilon \left( \eta_T + (u\eta)_{\xi} \right) = 0, \label{Bouss1DTxi1} \\
&\eta_{\xi} - u_{\xi} + \epsilon \left( u_T + uu_{\xi} + \frac 13 u_{\xi\xi\xi} \right) - \frac{\epsilon^2}{3}u_{T\xi\xi} = 0. \label{Bouss1DTxi2}
\end{align}
We seek a solution of (\ref{Bouss1DTxi1}) and (\ref{Bouss1DTxi2}) in the form of the asymptotic multiple scale expansion
\begin{equation}
\eta(T,\xi) = \eta^{(0)}(T,\xi) + \epsilon \eta^{(1)}(T,\xi) + \epsilon^2 \eta^{(2)}(T,\xi) + O(\epsilon^3),
\end{equation}
and similarly for $u$. To leading order, $O(1)$, this yields
\begin{equation}
u_{\xi}^{(0)} = \eta_{\xi}^{(0)}, \label{LOu}
\end{equation}
and hence $u^{(0)} = \eta^{(0)}$ provided the wave propagates into an unperturbed medium. At the second order, $O(\epsilon)$, we obtain
\begin{align}
&\eta_{\xi}^{(1)} - u_{\xi}^{(1)} = u_T^{(0)} + \left(u^{(0)} \eta^{(0)} \right)_{\xi}, \label{oe1} \\
&u_{\xi}^{(1)} - \eta_{\xi}^{(1)} = \eta_T^{(0)} + u^{(0)} u^{(0)}_{\xi} + \frac 13 u_{\xi\xi\xi}^{(0)}, \label{oe2}
\end{align}
from which taking the sum of (\ref{oe1}) and (\ref{oe2}), and substituting the leading order relation $u^{(0)} = \eta^{(0)}$ again yields the KdV equation
\begin{equation}
\eta_T^{(0)} + \frac 32 \eta^{(0)}\eta^{(0)}_{\xi} + \frac 16 \eta^{(0)}_{\xi\xi\xi} = 0. \label{kdva}
\end{equation}
At the third order, $O(\epsilon^2)$, we obtain
\begin{align}
&\eta_{\xi}^{(2)} - u_{\xi}^{(2)} = u_T^{(1)} + \left( u^{(1)} \eta^{(0)} \right)_{\xi} + \left( u^{(0)} \eta^{(1)} \right)_{\xi}, \label{oes1} \\
&u_{\xi}^{(2)} - \eta_{\xi}^{(2)} = \eta^{(1)}_T + \left( u^{(0)} u^{(1)} \right)_{\xi} + \frac 13 u^{(1)}_{\xi\xi\xi} - \frac 13 u^{(0)}_{T\xi\xi}, \label{oes2}
\end{align}
from which taking the sum of (\ref{oes1}) and (\ref{oes2}), and using the relations (\ref{expression1}), (\ref{expression2}), and (\ref{expression3}) yields
\begin{equation}
2 \eta_T^{(1)} + 3 \left( \eta^{(0)} \eta^{(1)} \right)_{\xi} + \frac 13 \eta^{(1)}_{\xi\xi\xi} - \frac{3}{4} \left. \eta^{(0)} \right.^2 \eta^{(0)}_{\xi} + \frac{7}{12} \eta^{(0)}_{\xi} \eta^{(0)}_{\xi\xi} + \frac 12 \eta^{(0)} \eta^{(0)}_{\xi\xi\xi} + \frac{1}{12} \eta^{(0)}_{\xi\xi\xi\xi\xi} = 0. \label{nearlyeKdVa}
\end{equation}
Approximating $\eta$ by $\eta \simeq \eta^{(0)} + \epsilon \eta^{(1)}$ and taking the sum of (\ref{kdva}) and $\epsilon$ lots of (\ref{nearlyeKdVa}) yields the eKdV equation from the 1D BP equations as
\begin{equation}
\eta_T + \alpha \eta \eta_{\xi} + \beta \eta_{\xi\xi\xi} + \epsilon \left( \alpha_1 \eta^2 \eta_{\xi} + \gamma_1 \eta \eta_{\xi\xi\xi} + \gamma_2 \eta_{\xi} \eta_{\xi\xi} + \beta_1 \eta_{\xi\xi\xi\xi\xi} \right) = 0, \label{eKdVa}
\end{equation}
where we have again truncated $O(\epsilon^2)$ terms and the coefficients $(\alpha, \beta, \alpha_1, \gamma_1, \gamma_2, \beta_1 )$ take the values $(3/2, 1/6, -3/8, 1/4, 7/24, 1/24)$. If we apply the change of variables $X = \epsilon \xi + T$ to (\ref{eKdVa}) we obtain the slow space variation as
\begin{equation}
\eta_X + \alpha \eta \eta_{\xi} + \beta \eta_{\xi\xi\xi} + \epsilon \left( \alpha_1 \eta^2 \eta_{\xi} + \gamma_1 \eta \eta_{\xi\xi\xi} + \gamma_2 \eta_{\xi} \eta_{\xi\xi} + \beta_1 \eta_{\xi\xi\xi\xi\xi} \right) = 0, \label{eKdVXa}
\end{equation}
where the coefficients $(\alpha, \beta, \alpha_1, \gamma_1, \gamma_2, \beta_1 )$ take the values $(3/2, 1/6, -21/8, -3/4, -47/24,\\ -1/24)$. The coefficients for the derivations given here are given in table \ref{table_ekdv_coeffs}.

\section{Numerical Schemes}
\label{appendixB}

Numerical solutions to all models discussed in this study are sought in the same manner. The temporal derivatives are approximated via the $4^{\text{th}}$-order Runge-Kutta (RK4) scheme and the spatial derivatives are approximated using an efficient pseudospectral scheme. The spacial accuracy of solution can be evaluated by ensuring there is sufficient decay in the Fourier coefficients, see for example \citet{KR11,DK22}, and for the 1D {\crr Serre} equations we also observe the relevant conserved quantities.

We discretise the spacial domain by $\Delta_{\xi} = (\xi_{\text{max}}-\xi_{\text{min}})/N_{\xi}$ such that $\xi_i = \xi_{\text{min}} + i \Delta_{\xi}$, for $i = 0,1,\ldots, N_{\xi}$ and similarly the temporal discretisation used is $\Delta_t = (t_{\text{max}}-t_{\text{min}})/N_t$ such that $t_n = t_{\text{min}} + n \Delta_t$, for $n = 0,1,\ldots, N_t$. The standard RK4 scheme is employed for $\eta_t = f(t,\xi)$ such that
\begin{align}
&\eta^{n+1} = \eta^n + \frac{\Delta_t}{6} \left( k_1 + 2k_2 + 2k_3 + k_4 \right), \\
&k_1 = f\left( t_n, \eta^n \right), \quad k_2 = f\left( t_n + \frac{\Delta_t}{2}, \eta^n + \frac{k_1 \Delta_t}{2} \right), \\
&k_3 = f\left( t_n + \frac{\Delta_t}{2}, \eta^n + \frac{k_2 \Delta_t}{2} \right), \quad k_4 = f\left( t_n + \Delta_t, \eta^n + k_3 \Delta_t \right).
\end{align}
Partial derivatives of the discretised field $\eta$ are approximated as
\begin{equation}
\frac{\partial^n \eta}{\partial \xi^n} = \mathcal{F}^{-1}\left[ (ik)^n \mathcal{F}[\eta] \right],
\end{equation}
where $k$ are the wavenumbers, and $\mathcal{F}$ and $\mathcal{F}^{-1}$ are the discrete Fourier and inverse discrete Fourier transform, respectively \citep{T00}. For derivatives of order two or more we truncate the wavenumbers via Orszag's $2/3$ rule to remove de-aliasing errors \citep{O71}, however, no further filtering is applied to preserve the resonant radiation of interest in this study.

Numerical solutions of the 1D SGN equations are sought in $(t,\xi)$-variables where we take the temporal step size $\Delta_t = 10^{-3}$ and spacial step size $\Delta_{\xi} = 10^{-1}$. We therefore compute solutions for $t \in[0, T_{\text{max}}/ \epsilon]$ to match the reduced order models final time. In $(t,\xi)$-variables it is possible to rewrite the 1D SGN equations (\ref{SGN1D1}) and (\ref{SGN1D2}) as
\begin{align}
&\frac{\partial \eta}{\partial t} = \frac{\partial}{\partial \xi} \left( \eta - (1 + \epsilon \eta)u \right), \label{sgnApp1} \\
&\frac{\partial q}{\partial t}    = \frac{\partial}{\partial \xi} \left( q - \eta - \epsilon u q + \frac{\epsilon}{2} u^2 + \frac{\epsilon\delta^2}{2}(1 + \epsilon\eta)^2 u_{\xi}^2 \right), \label{sgnApp2} \\
&q = u - \frac{\delta^2}{3(1 + \epsilon \eta)} \left( (1 + \epsilon\eta)^3 u_{\xi} \right)_{\xi}, \label{sgnApp3}
\end{align}
where (\ref{sgnApp1}) has been used to simplify the expression for (\ref{sgnApp2}) \citep{DCMM13,DK22}. To iterate (\ref{sgnApp1}) and (\ref{sgnApp2}) in time it is necessary to simultaneously invert the elliptic ODE, (\ref{sgnApp3}), for $u$. To do so we use the method outlined by \citet{DK22}. In Fourier space (\ref{sgnApp3}) takes the form $\hat M \hat u = \hat q$ where $\hat M$ is a nonlinear differential operator and hence its product is a convolution. We precondition this system such that $P^{-1} Mu = P^{-1} q$ where $\hat P = 1 + \delta^2 k^2/3$ and use a Krylov subspace iterative GMRES method \citep{SS86} to determine $u$. In practice the solution converges in a few iterations with the residual error stopping criteria of $5 \times 10^{-16}$, {\crr however, the computation time is still significantly greater than that of the KdV-type models also solved in this study due to inverting this operator at each time step.} For larger $\epsilon$ and $\delta$ values the number of iterations rises due to the linear preconditioner diverging from $\hat P^{-1} \hat M \simeq I$. This is easily counteracted by smaller time steps where $\hat u$ will only minimally change in each time-step iteration.

\begin{figure}
	\centerline{\includegraphics[width=0.90\linewidth]{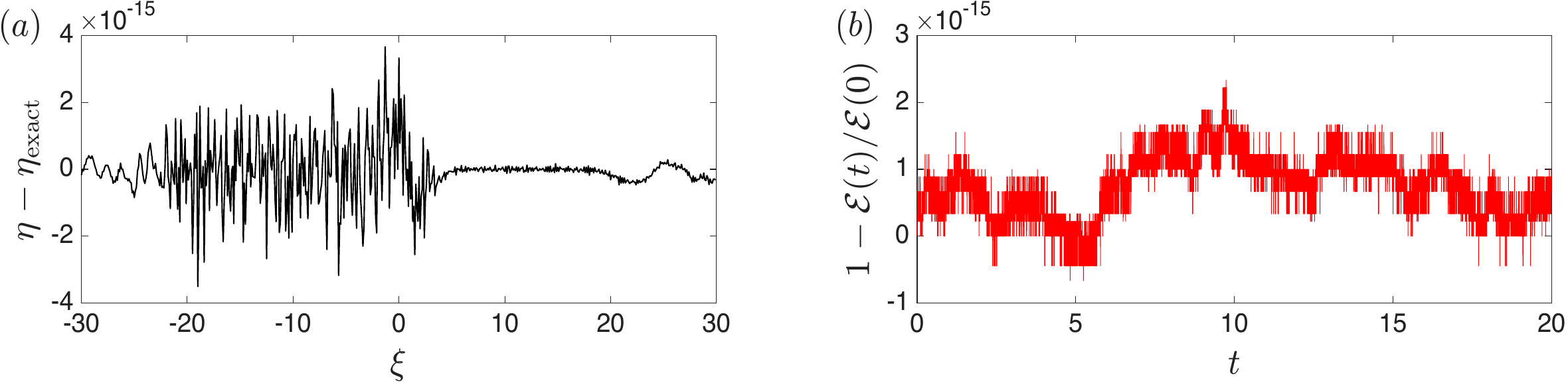}}
	\caption{Computation results of the 1D {\crr Serre} equations (\ref{sgnApp1}), (\ref{sgnApp2}), and (\ref{sgnApp3}) where (\textit{a}) is the error in $\eta$ at the final time $t = 20$, and (\textit{b}) the conservation of energy, $\mathcal{E}$, where $\mathcal{E}(0)$ is the energy computed at $t = 0$.}
	\label{sgnconservation}
\end{figure}

The numerical solver of the {\crr Serre} system (\ref{sgnApp1}), (\ref{sgnApp2}), and (\ref{sgnApp3}) is tested using the exact soliton solution, given by for example \citet{MDC17,DK22}, such that
\begin{equation}
\eta(t,\xi) = v \operatorname{sech}^2 \left( \sqrt{\frac{3\epsilon v}{4(1 + \epsilon v)}} \frac{1}{\delta} \left(\xi - (c-1)t \right) \right), \qquad u(t,\xi) = \frac{c \eta(t,\xi)}{1 + \epsilon \eta(t,\xi)}, \label{sgnsol}
\end{equation}
where $c = \sqrt{1 + \epsilon v}$ and $v$ is a free parameter determining the wave amplitude. We also monitor the known conserved quantities mass, momentum, energy, and the scaled tangential fluid velocity, $q$ \citep{DK22}. The energy of the 1D {\crr Serre} system  can be given in $(t,\xi)$-variables as
\begin{equation}
\frac{\mathrm{d}\mathcal{E}}{\mathrm{d}t} = \frac{\epsilon}{2} \int_{\mathcal{D}} (1 + \epsilon \eta) u^2 + \frac{\delta^2}{3} (1 + \epsilon \eta)^3 u_{\xi}^2 + \eta^2 ~ \mathrm{d}\xi = 0,
\end{equation}
where $\mathcal{D}$ is the entire $\xi$-domain \citep{MDC17}. An example of the conservation of energy and error in the computed solution is shown in (\ref{sgnconservation}) for $\epsilon = 0.1$, $\xi \in [-30,30]$, and $t \in [0,20]$.

The error in both $\eta$ and $\mathcal{E}(t)$ is $O(10^{-15})$ and close to machine double precision which is considerably below plotting accuracy. There is also sufficient decay in the Fourier coefficients, not pictured here, which indicates that no further spatial resolution is obtained from an increase in collocation points.

The final equations to solve are the reduced amplitude models and they are sought in the form given by \citet{T00}. As noted in Section \ref{section_disp_rel}, all of the reduced models studied here can be written as a combination of linear and nonlinear parts in the form $\eta_T + \mathcal{L}[\eta] + \mathcal{N}[\eta] = 0$. Therefore all of the reduced order models can be written in Fourier space as
\begin{equation}
\hat \eta_T + i \omega(k) \hat \eta = - \alpha \hat\eta \hat\eta_{\xi} - \epsilon \left( \alpha_1 \hat\eta^2 \hat\eta_{\xi} - \gamma_1 \hat\eta \hat\eta_{\xi\xi\xi} - \gamma_2 \hat\eta_{\xi} \hat\eta_{\xi\xi} \right), \label{numericalekdv}
\end{equation}
where $\omega(k)$ is the relevant linear dispersion relation and the coefficients also take the required values. For the reduced order models it is important to remove the stiff terms, encapsulated by $i \omega(k) \hat \eta$, improving the stability properties. In turn this increases the minimum time step and reduces the computation time considerably. See for example the studies by \citet{STCK24,STCK25}. To do so we introduce the integrating factor $\Lambda$ such that $\hat q = \Lambda \hat \eta$ where
\begin{equation}
\Lambda = \exp\left( i \omega(k) T \right).
\end{equation}
Upon expanding the $T$-derivative in (\ref{numericalekdv}) this yields
\begin{equation}
\hat q_T = - \Lambda \left( \alpha \hat \eta \hat \eta_{\xi} + \epsilon \left(  \alpha_1 \hat{\eta}^2 \hat \eta_{\xi} + \gamma_1 \hat \eta \hat \eta_{\xi\xi\xi} + \gamma_2 \hat \eta_{\xi} \hat \eta_{\xi\xi} \right) \right). \label{numericq}
\end{equation}
{\crr Exploiting the idea of \citet{T00} we can eliminate $\hat q$ and the numerically detrimental exponential dependence upon $T$. Taking the right hand side of (\ref{numericalekdv}) to be $f(T,\hat\eta)$ we can rewrite the RK4 scheme as
\begin{align}
&\hat \eta^{n+1} = E^2 \hat \eta^n + \frac{\Delta_T}{6} \left( k_1 + 2k_2 + 2 k_3 + k_4 \right), \\
&k_1 = E^2 ~ f\left( T_n, \hat\eta^n \right), \quad k_2 = E ~ f\left( T_n + \frac{\Delta_T}{2}, E ~ \hat\eta^n + \frac{k_1 \Delta_t}{2E} \right), \\
&k_3 = E ~ f\left( T_n + \frac{\Delta_T}{2}, E ~ \hat\eta^n + \frac{k_2 \Delta_T}{2E} \right), \quad k_4 = f\left( T_n + \Delta_T, E^2 ~ \hat\eta^n + k_3 \Delta_T \right),
\end{align}
where $E = \exp(- i\omega \Delta_T / 2)$. Numerical solutions are efficiently sought as there is no requirement to invert the integrating factor, and as $T$ increases the component $E$ remains small which reduces rounding errors. We can also increase $\Delta_T$ from $\Delta_T = 10^{-6}$ to $\Delta_T = 5 \times 10^{-4}$ compared to the original stiff problem with explicit high order dispersive terms. Multiplication in $f$ is done in real space and then returned to Fourier space for time stepping and calculating derivatives.}

For certain simulations it is convenient to suppress small waves that propagate towards the boundaries, and since the pseudospectral scheme is periodic, these features would propagate back into the domain and interfere with the simulation. In these cases we introduce a sponge layer by mapping $\eta_T \rightarrow \eta_T + s_L \eta$, and similarly for $u$ and $q$, where we take the sponge function to be
\begin{equation}
s_L(\xi) = \sigma \left(1 - \frac 12 \bigl( \tanh \kappa ( \xi - \xi_{\text{min}} - \xi_{\text{span}} ) - \tanh \kappa ( \xi - \xi_{\text{max}} + \xi_{\text{span}} ) \bigr) \right),
\end{equation}
and typically take $\xi_{\text{span}} = (\xi_{\text{min}} - \xi_{\text{max}} ) / 20$. The sponge layer acts as zero in the central portion of the domain and increases near the boundary at a point determined by $\xi_{\text{span}}$ that dampens waves as they propagate into this region. The parameters $\sigma = 750$ and $\kappa = 1$ {\crr did not} produce any reflected waves in the experiments.


\end{document}